\documentclass{article}
\usepackage[utf8]{inputenc}
\usepackage{authblk}
\usepackage{setspace}
\usepackage[margin=1.25in]{geometry}
\usepackage{graphicx}
\usepackage{subcaption}
\usepackage{lineno}
\usepackage{url}            
\usepackage{lastpage}
\usepackage{amsmath, amssymb, amsthm, amsfonts}
\usepackage{color}
\usepackage{caption}
\usepackage{mathtools}
\usepackage[algo2e]{algorithm2e} 
\usepackage{mathrsfs} 
\usepackage{algorithm}
\usepackage{natbib}
\usepackage{enumitem}
\usepackage[colorlinks=true, urlcolor=blue, linkcolor=blue, citecolor=blue]{hyperref}

\DeclareMathOperator{\yvec}{y}
\DeclareMathOperator*{\ff}{\boldsymbol{f}}
\DeclareMathOperator{\ymat}{Y}
\DeclareMathOperator{\xvec}{x}
\DeclareMathOperator{\avec}{a}
\DeclareMathOperator{\xproc}{X}

\DeclareMathOperator*{\var}{var}
\DeclareMathOperator*{\diag}{diag}
\DeclareMathOperator*{\argmax}{argmax}

\newtheorem{prop}{Proposition}
\newtheorem{lemma}{Lemma}
\providecommand{\keywords}[1]{\textbf{\textit{Keywords:}} #1}

\title{\textbf{Functional Generalized Canonical Correlation Analysis for studying multiple longitudinal variables}}

\author[1]{Lucas Sort}
\author[1]{Laurent Le Brusquet}
\author[1]{Arthur Tenenhaus}

\affil[1]{\textit{Université Paris-Saclay, CNRS, CentraleSupélec, Laboratoire des Signaux et Systèmes, Gif-sur-Yvette, France}}
\affil[ ]{\textit{\texttt{lucas.sort@centralesupelec.fr}}}

\date{}

\begin{document}

\maketitle

\begin{abstract}
In this paper, we introduce Functional Generalized Canonical Correlation Analysis (FGCCA), a new framework for exploring associations between multiple random processes observed jointly. The framework is based on the multiblock Regularized Generalized Canonical Correlation Analysis (RGCCA) framework. It is robust to sparsely and irregularly observed data, making it applicable in many settings. We establish the monotonic property of the solving procedure and introduce a Bayesian approach for estimating canonical components. We propose an extension of the framework that allows the integration of a univariate or multivariate response into the analysis, paving the way for predictive applications. We evaluate the method's efficiency in simulation studies and present a use case on a longitudinal dataset.
\end{abstract}


\newpage

\section{Introduction}
\label{sec1}

Measuring multiple biomarkers jointly over time is common in observational studies and clinical trials. As they characterize various biological processes which are often interdependent, those biomarkers are usually correlated. Hence, separately analyzing those longitudinal variables may hide parts of the biological mechanisms at stake and give redundant information. Furthermore, as subjects often miss one or more visits, the biomarkers may be observed sparsely and irregularly. Therefore, along with the complex time-dependent continuous structure of the data, the statistical analysis of multiple longitudinal variables requires using specific methodologies for efficiently harvesting information and integrating the interaction between the variables.

In the multivariate setting, the analysis of data coming from multiple sources, usually represented by multiple sets of variables, is often referred to as "multiblock" or "multi-set" analysis. Canonical Correlation Analysis (CCA) (\citet{Hotelling1936}) is one of the most notorious approaches, but it is limited to exploring associations between only two sets of variables. Therefore, various methods were proposed to generalize the CCA problem to more than two sets of variables (\citet{Horst1961, carroll1968generalization, KETTENRING1971}). More recently, \citet{Tenenhaus2011} introduced Regularized Generalized Canonical Correlation Analysis (RGCCA), a regularized and more flexible framework for studying multiple sets of variables, giving birth to a new generation of methods (\citet{Tenenhaus2014, Tenenhaus2015, Singh2019}). Various extensions of RGCCA were proposed to handle emerging data types, such as tensor data (\citet{GIRKA}). However, most are still limited to finite-dimensional Euclidian spaces and are not designed to handle large amounts of missing observations.

In the longitudinal literature, approaches based on linear mixed-effects models have been widely employed over the past decades for studying longitudinal biomarkers (\citet{Rizopoulos2011, German2021}). However, for adapting multivariate data analysis methods to the longitudinal setting, functional approaches have been preferred as functional spaces are easy to handle and allow to describe the underlying smooth structure of the time-dependent variables. In this context, adaptations of Principal Component Analysis (PCA) to the longitudinal setting have flourished (\citet{Rice1991}). Most notably, \citet{Yao2005} proposed an adaptation using a covariance-based procedure and a Bayesian approach to estimate the principal components. The method is thus robust to sparse and irregular data, making it applicable to numerous problems. 

Multiple extensions of CCA were proposed to explore associations between two longitudinal variables (\citet{Leurgans1993, He2003, Zhou2008, Shin2015}). Regularization is crucial in this infinite-dimensional context as CCA requires inverting covariance matrices. Inspired by \citet{Yao2005} approach, \citet{Yang2011} introduced the Functional Singular Value Decomposition (FSVD), which moves the CCA criterion to a covariance criterion and also uses a Bayesian approach to estimate the canonical components. However, as in the multivariate setting, those adaptations are limited to a pair of longitudinal variables. Few methods can go beyond this limitation. To our knowledge, \citet{Hwang2011} proposed the first approach to find associations between any number of longitudinal variables using a homogenous components criterion. More recently, \citet{Gorecki2020} proposed to adapt \citet{Horst1961} approach to functional spaces using basis decomposition. Although it is not designed to explore association among several longitudinal variables, it is worth mentioning Multivariate Functional Principal Component Analysis (MFPCA) (\citet{Happ2018}), designed to retrieve the principal modes of variation on multivariate longitudinal data. The method can also handle sparse and irregular data.

In this context, we propose Functional Generalized Canonical Correlation Analysis (FGCCA), a framework based on RGCCA that allows exploring and studying associations between several longitudinal markers in a flexible way. The method proposed is robust to sparse and irregular longitudinal data. Furthermore, it is designed so it can integrate a multivariate block in the analysis to perform, for instance, supervised learning. As RGCCA, the framework provided by FGCCA is so vast that it encompasses many existing methods, notably \citet{Yao2005} FPCA, FSVD, or Functional Partial Least Squares (FPLS) as presented by \citet{Preda2007}.

The paper is organized as follows. First, in Section \ref{sec:rgcca}, we recall the Regularized Generalized Canonical Correlation Analysis (RGCCA) framework. Then, in Section \ref{sec:fgcca}, we introduce the Functional Generalized Canonical Correlation Analysis (FGCCA) context and optimization problem. We present the solving procedure of FGCCA in Section \ref{sec:resolution} and introduce a scheme for retrieving higher-order functions and estimating components in Section \ref{sec:higherorder} and \ref{sec:components} respectively. We validate our method on simulation studies in Section \ref{sec:simulation} and propose an application to a real dataset in Section \ref{sec:application}. Finally, our approach's limitations and possible extensions are discussed in Section \ref{sec:dicussion}.

Proofs of propositions are given in Supplementary Materials. The code used to run the experiments and the R implementation of FGCCA are freely available on github: \url{https://github.com/Sort-L/FGCCA-Code}.

\section{Method}
\label{sec:methods}

\subsection{Regularized Generalized Canonical Correlation Analysis (RGCCA)}
\label{sec:rgcca}

Regularized Generalized Canonical Correlation Analysis (RGCCA) (\cite{Tenenhaus2017}) is an optimization and statistical framework for studying associations between multiple sets of random variables. Denoting $\xvec_1, \dots, \xvec_J$ the sets of respectively $p_1, \dots, p_J$ random variables, and $\boldsymbol{\Sigma}_{jj'}$ the $p_j \times p_{j'}$ matrix of (cross-)covariance between $\xvec_j$ and $\xvec_{j'}$, the RGCCA optimization problem can be expressed as:
\begin{equation}
\label{eq:rgcca}
        \argmax_{\avec_1, \dots, \avec_J \in \Omega_1\times\dots\times\Omega_J} \sum_{j \neq j'} c_{j,j'}g( \avec_j^\top \boldsymbol{\Sigma}_{jj'} \avec_{j'}) \\
\end{equation}
where $\Omega_j$ is defined as $\Omega_j = \left\{ \avec_j \in \mathbb{R}^{p_j}  \mid  \avec_j^\top \boldsymbol{M}_j \avec_j = 1 \right\}$, with $\boldsymbol{M}_j$ being a symmetric-positive matrix, $g$ is a convex differentiable function and the matrix $C=(c_{j,j'})$ is a $J\times J$ symmetric matrix with positive elements specifying the desired connection design to study associations between the blocks. Classically we set $c_{j,j'} =  1$ if we want to consider the interaction between the blocks $j$ and $j'$ and $c_{j,j'} = 0$ otherwise. Additionally, we often consider $\boldsymbol{M}_j = \tau_j \boldsymbol{I}_{p_j} + (1-\tau_j) \boldsymbol{\Sigma}_{jj}$ with $\tau_j \in [0,1]$  to interpolate smoothly the criterion between a correlation criterion, when $\tau_j=0$, and a covariance criterion, when $\tau_j=1$. In this context, the goal of RGCCA is to retrieve block weight vectors $\avec_j$ giving block components $\yvec_j = \xvec_j^\top \avec_j$, which are a compromise of the information from each set of variables and the information shared with the other sets of variables.


Finally, two strategies are often used to compute higher-level weight vectors for each set of variables. The first strategy leads to new weight vectors associated with components uncorrelated to the previous ones. The second strategy yields new weight vectors orthogonal to the previous ones. Both strategies require transforming the original sets of variables $\xvec_j$ into new sets of variables $\xvec_j'$ called "deflated" vectors. The first transformation, more often used, consists in regressing out from each set $\xvec_j$ its associated component $\yvec_j = \xvec_j^\top a_j$: the transformation can be written $ \xvec_j' = \xvec_j-(a_j^\top \boldsymbol{\Sigma}_{jj} a_j)^{-1} \boldsymbol{\Sigma}_{jj} a_j a_j^\top \xvec_j $. The second transformation consists in projecting each set $\xvec_j$ onto the orthogonal of the space spanned by the previous weight vectors: it is defined as $\xvec_j' = \xvec_j - \avec_j \avec_j^{\top}\xvec_j$. To retrieve new weight vectors and new components with the desired properties, the solving procedure is rerun, replacing original vectors $\xvec_j$ with deflated vectors $\xvec_j'$. The transformation equations, often called "deflation" equations, can be used repeatedly to retrieve multiple weight vectors and components.

As demonstrated in \citet{Tenenhaus2017}, the framework of RGCCA is very general and subsumes many notorious data analysis methods such as Principal Component Analysis (PCA), Canonical Correlation Analysis (CCA) (\citet{Hotelling1936}), Partial Least Squares (PLS) regression (\citet{Wold2001}), and Generalized Canonical Correlation Analysis (GCCA) (\cite{carroll1968generalization}, \citet{Horst1961}, \citet{KETTENRING1971}), to name a few. Many extensions and adaptations have been proposed to tackle a wide variety of problems, but, to our knowledge, none exists to integrate the time continuous structure of longitudinal data and, especially, to handle highly sparse and irregular observations.

\subsection{Functional Generalized Canonical Correlation Analysis (FGCCA)}
\label{sec:fgcca}

We now consider multiple time-dependent variables, such as longitudinal biomarkers. We propose to adapt the previous framework to functional spaces, and more precisely square-integrable random processes, since random processes can represent the time continuous structure of the data. Our approach, named Functional Generalized Canonical Correlation Analysis (FGCCA) is now introduced.

\subsubsection{Notations and definitions}

From now on, we consider $\xproc_1, \dots, \xproc_J$, $J$ square-integrable random processes defined on compact intervals of $\mathbb{R}$, $I_1, \dots, I_J$ respectively. Note that the random objects are thus part of infinite-dimensional Hilbert spaces $L^2(I_1),\dots, L^2(I_J)$. In this context, we define for the process $j$ the mean function $\mu_j$ as $\mu_j(t) = \mathbb{E}(\xproc_j(t))$ for $t \in I_j$. Additionally, the (cross-)covariance function (or "surface") $\Sigma_{jj'}$ between the processes $j$ and $j'$ is defined as $\Sigma_{jj'}(s, t) = \mathbb{E}((\xproc_j(s) - \mu_j(s))(\xproc_{j'}(t) - \mu_{j'}(t)))$ with $s \in I_j$ and $t \in I_{j'}$. Finally, the (cross-)covariance operator between the processes $j$ and $j'$, $\boldsymbol{\Sigma}_{jj'}$, is defined as :
\begin{align*}
    \boldsymbol{\Sigma}_{jj'} &: L^2(I_{j'}) \rightarrow L^2(I_j), \; f \mapsto g, \; g(s) = \int_{I_{j'}} \Sigma_{jj'}(s, t)f(t)\text{dt}
\end{align*}

\subsubsection{Model}

Following the optimization problem (\ref{eq:rgcca}), defining RGCCA, we define Functional Generalized Canonical Correlation Analysis (FGCCA) optimization problem by moving from the multivariate setting of $\mathbb{R}^{p_j}$ spaces to the functional setting of $L^2(I_j)$ spaces, replacing the sets of variables $\xvec_j$ by the random processes $\xproc_j$ and, therefore, the euclidean dot product $a^{\top}b$ by the functional scalar product defined by $\langle f, g \rangle_{L^2} = \int fg$. The FGCCA optimization problem can therefore be written:
\begin{equation}
\label{eq:fgccacv}
        \argmax_{f_1, \dots, f_J \in \Omega_1\times\dots\times\Omega_J} \sum_{j \neq j'} c_{j,j'}g(\langle f_j, \boldsymbol{\Sigma}_{jj'} f_{j'}\rangle_{L^2})
\end{equation}
where $\Omega_j$ is defined as $\Omega_j = \left\{ f_j \in L^2(I_j)  \mid  \langle f_j, \boldsymbol{M}_j f_j \rangle_{L^2} = 1 \right\}$ with $\boldsymbol{M}_j$ being a symmetric positive-definite operator, and where $g$ and $\bold{C}=(c_{j,j'})$ are defined similarly as before. Similarly to RGCCA, we suggest setting $\boldsymbol{M}_j = \tau_j \boldsymbol{I}_{I_j} + (1-\tau_j) \boldsymbol{\Sigma}_{jj}$. However, to ensure the positive definiteness of the operator $\boldsymbol{M}_j$, regularization parameters $\tau_j$ must be strictly superior to $0$ (and thus, lie in $]0, 1]$) as covariance operators $\boldsymbol{\Sigma}_{jj}$ are not necessarily definite in the infinite-dimensional setting. Functions $f_j$ and components $\yvec_j = \langle \xproc_j, f_j, \rangle_{L^2}$, allow capturing information for each process which, depending on the model parameters, is a summary of both the information from each process and the information shared with the others. 

\subsubsection{(Cross-)Covariance estimation} In the multivariate setting, the most straightforward estimation for (cross-)covariance matrices $\boldsymbol{\Sigma}_{jj'}$ is the sample covariance matrix, which is fast and easy to compute. However, in the functional setting, it is usually preferable to use alternative strategies that integrate the data's time-continuous structure.

Various methods have been proposed over the past decades to integrate this structure. In the functional data analysis literature, 
(cross-)covariance operators $\boldsymbol{\Sigma}_{jj'}$ are often discretized and estimated on dense \& regular grids using kernel smoothing methods (\citet{Yao2005}, \citet{Yang2011}) or Generative Additive Models (GAM), which are easy to implement. In this context, estimation methods often have several hyperparameters that must be set. Usually, those hyperparameters are manually specified prior to the analysis. However, cross-validation procedures, such as leave-one-out cross-validation or criterion-based procedures, can be used for selecting them (\citet{Leurgans1993}). Additionally, due to approximation errors, estimated operators are rarely positive in practice, making the choice of the regularization parameters $\tau_j$ in FGCCA crucial, as the interval they are defined on may not be clearly identified. Consequently, we advise setting regularization parameters to $1$ as it will always prevent the optimization problem from being ill-posed.

Finally, we recommend normalizing the different processes before estimating (cross-)covariance operators. Like in the multivariate setting, considerable differences in the variance of the processes may lead to biased results. For this purpose, we suggest using the normalization presented in a similar context by \citet{Happ2018}, enforcing the integrated variance to be the same for each process. To achieve this, each process $j$ is multiplied by the following normalization quantity: 
\begin{equation*}
    w_j = \left ( \int_{{I_j}}\var(\xproc_j(t))\text{dt} \right )^{-1/2}
\end{equation*}

\subsection{Resolution}
\label{sec:resolution}

We now introduce a procedure to retrieve solutions to the FGCCA optimization problem. Convergence properties are given, ensuring the stability of the solving procedure.

\subsubsection{Procedure}
\label{sec:solving}
 Let $\Psi$ be the objective function:
\begin{equation*}
    \Psi(f_1, \dots, f_J) = \Psi(\boldsymbol{f}) = \sum_{j \neq j'} c_{j,j'}g(\langle f_j, \boldsymbol{\Sigma}_{jj'} f_{j'}\rangle_{L^2})
\end{equation*}
As $\Psi$ is a function of multiple arguments, we suggest using a block coordinate ascent (BCA) strategy (\cite{DeLeeuw1994}) for finding solutions to the maximization problem. This strategy consists in maximizing $\Psi$ argument by argument until convergence is reached. The properties of $g$ implies that the objective function is differentiable and multi-convex, meaning that it is convex with respect to each argument $f_j$ when all the others are fixed. Note also that we consider here the "functional differentiability" since $\Psi$ has functional arguments. A proof of the definition of the gradient is given in Supplementary Materials. From these properties we can derive the following inequality for $\tilde{f}_j \in \Omega_j$:
\begin{align}
    \Psi(f_1, \dots, \tilde{f}_j, &\dots, f_J) \geq \Psi(\boldsymbol{f}) + \langle\nabla_j \Psi(\boldsymbol{f}),\tilde{f}_j - f_j\rangle = m_j(\boldsymbol{f}, \tilde{f}_j) 
\end{align}
where $\nabla_j \Psi(\boldsymbol{f})$ is the functional partial derivative of $\Psi$ with respect to the $j$th function. With this expression, we notice that maximizing $\Psi$ for the $j$th argument can be achieved by maximizing the minorizing function $m_j$. In this minorizing function, only the term $\langle\nabla_j \Psi(\boldsymbol{f}) ,\tilde{f}_j\rangle$ is relevant since all the others are fixed. Therefore, the maximum of $m_j$ under the constraint that $\tilde{f}_j \in \Omega_j$ is reached for:
\begin{equation}
    \hat{f}_j = \argmax_{\tilde{f}_j\in\Omega_j} \langle\nabla_j \Psi(\boldsymbol{f}), \tilde{f}_j\rangle = \frac{\boldsymbol{M}_j^{-1} \nabla_j \Psi(\boldsymbol{f})}{||\boldsymbol{M}_j^{-1/2}\nabla_j \Psi(\boldsymbol{f})||} := r_j(\boldsymbol{f})
\end{equation}
where the partial derivative can be expressed as:
\begin{equation}
\label{eq:grad}
    \nabla_j \Psi(\boldsymbol{f}) = 2\sum_{\substack{j'=1 \\ j'\neq j}}^J c_{j, j'} g'(\langle f_j, \boldsymbol{\Sigma}_{jj'}f_{j'}\rangle)\boldsymbol{\Sigma}_{jj'}f_{j'}
\end{equation}
From this, we propose the Algorithm (\ref{alg:fgcca}) for retrieving solutions to optimization problem (\ref{eq:fgccacv}). 

\begin{algorithm}
\caption{FGCCA algorithm}
\label{alg:fgcca}
\KwData{$(\boldsymbol{\Sigma}_{jj'})_{1 \leq j, j' \leq J}, \, g, C, \epsilon, \boldsymbol{f}^{0}$}
\KwResult{$f_1^{s+1},\dots,f_J^{s+1}$}
\Repeat{$\Psi(f_1^{s+1},\dots,f_J^{s+1}) - \Psi(f_1^{s},\dots,f_J^{s}) < \epsilon$}{
    \For{$j=1$ \KwTo $J$}{
        \begin{equation*}
            \tilde{f}_j^{s+1} = \frac{\boldsymbol{M}_j^{-1} \nabla_l \Psi(f_1^{s+1},\dots,f_{j-1}^{s+1}, f_j^s, f_{j+1}^s, \dots, f_J^s)}{||\boldsymbol{M}_j^{-1/2} \nabla_l \Psi(f_1^{s+1},\dots,f_{j-1}^{s+1}, f_j^s, f_{j+1}^s, \dots, f_J^s)||}
        \end{equation*}\;
    } 
    $s = s + 1$\;
}
\end{algorithm}

\subsubsection{Monotone convergence}

Denoting $\Omega = \Omega_1\times\dots\times\Omega_J$ we define $c_j : \Omega \rightarrow \Omega$ as the operator $c_j(\boldsymbol{f}) = (f_1,\dots; f_{j-1}, r_j(\boldsymbol{f}), f_{j+1},\dots, f_J)$ with $r_j(\boldsymbol{f})$ being the update function for the $j$th function of the solving procedure in Section \ref{sec:solving}. We also define $c : \Omega \rightarrow \Omega$ as the operator $c = c_J \circ \dots \circ c_1$

We consider the sequence $\{\boldsymbol{f}^s = (f_1^s,\dots, f_J^s)\}$ generated by $\boldsymbol{f}^{s+1} = c(\boldsymbol{f}^s)$. The following proposition states the monotone convergence of the generated sequence $\{\boldsymbol{f}^s\}^{\infty}_{s=0}$ and holds as long as the update $r_j(\boldsymbol{f})$ exists, is unique and $\Omega$ is bounded:

\begin{prop}
Considering any sequence $\{ \boldsymbol{f}^s \}_{s=0}^{\infty}$ generated recursively by the relation $\boldsymbol{f}^{s+1} = c(\boldsymbol{f}^s)$ with $\boldsymbol{f}^0 \in \Omega$. The sequence $\{\Psi(\boldsymbol{f}^s)\}$ is monotonically increasing and therefore convergent as $\Psi$ is bounded on $\Omega$, implying the convergence of the FGCCA algorithm.
\end{prop}

Using this proposition, since $\Omega$ is bounded and $r_j(\boldsymbol{f})$ properly and uniquely defined for all $j$ with the functional gradient, we can conclude that the Algorithm (\ref{alg:fgcca}) is monotone and convergent.

\subsection{Retrieving higher-order orthogonal functions}
\label{sec:higherorder}

Solving the optimization problem as described previously only yields one function per block. However, it is often preferable to retrieve multiple functions leading to multiple components. For this purpose, we suggest, as in the RGCCA framework, using a deflation strategy. 


First, we propose considering a deflation strategy for retrieving orthogonal vectors. As detailed in Section \ref{sec:rgcca}, the deflation equation associated with this strategy is $\xvec_j' = \xvec_j - \avec_j\avec_j^{\top} \xvec_j$. This expression may be unusable in the functional setting, with sparse and irregular observations, as processes $\xproc_j$ are not fully observed. Moreover, only the (cross-)covariance operators are involved in the solving procedure, making the deflation of the blocks appear as an unnecessary step in the procedure. Therefore, we establish the following proposition, allowing us to deflate the (cross-)covariance operators directly without involving the possibly ill-defined blocks:

\begin{prop}
    Denoting $\boldsymbol{\Sigma}_{jj'}'$ the deflated (cross-)covariance operator of $\boldsymbol{\Sigma}_{jj'}$, obtained after projecting the processes onto the orthogonal of the space spanned by their associated vectors. The following equality holds:
    \begin{align}
        \boldsymbol{\Sigma}_{jj'}' &= (\boldsymbol{I}_{I_j} - \boldsymbol{\Phi}_j)\boldsymbol{\Sigma}_{jj'}(\boldsymbol{I}_{I_{j'}} - \boldsymbol{\Phi}_{j'})
    \end{align}
    where $\boldsymbol{\Phi}_j : L^2(I_j) \rightarrow L^2(I_j)$ is the operator defined by :
    \begin{equation*}
        (\boldsymbol{\Phi}_j)(f) = (f_j \otimes f_j)(f) = \langle f_j, f\rangle f_j
    \end{equation*}
\end{prop}

To retrieve new orthogonal functions, deflated operators are plugged into the optimization problem and Algorithm \ref{alg:fgcca} is run again. The deflation procedure can be repeated multiple times, allowing to retrieve a set of canonical functions $\{f_j^m\}_{1 \leq m \leq M}$ for each random process. The number of canonical functions to retrieve is often manually set but it can be chosen using cross-validation or criterion-based approaches. Finally, in this context, the set of components $\{\yvec_j^m\}_{1 \leq m \leq M}$ obtained for each process can be directly estimated from the original processes $\xproc_j$ without using the deflated processes $\xproc_j^m$, a desirable property in our setting as it may be difficult to compute and manipulate the deflated processes. Indeed, we can demonstrate easily that $\langle \xproc_j, f_j^m \rangle_{L^2} = \langle \xproc_j^m, f_j^m \rangle_{L^2} = \yvec_j^m$, where $\xproc_j^m$ stands for the $m$th deflation of process $j$.

\subsection{Estimating components}
\label{sec:components}

Computing the components $\yvec_j^m = \langle \xproc_j, f_j^m \rangle_{L^2}$ may be difficult in the sparse and irregular setting, as the numerical estimation of the $L^2$ scalar product can be unstable and untracktable, particularly if the number of observations is small. In this context, inspired from \citet{Yao2005} and \citet{Yang2011}, we propose to estimate the components using a Bayesian approach.

\subsubsection{Notations}
In the following, subscripts $i$, $j$, $k$ denote respectively the subject number, the process number, and the observation number. We denote $n_{ij}$ the number of observations, $\bold{U}_{ij} = (U_{ij1},\dots, U_{ijn_{ij}})^{\top}\in\mathbb{R}^{n_{ij}\times 1}$ the observations, and $t_{ij} = (t_{ij1}, \dots, t_{ijn_{ij}})$ the observation time points. Finally the observations are modeled as:
\begin{equation}
    U_{ijk} = X_{ij}(t_{ijk}) + \varepsilon_{ijk}
\end{equation}
where $X_{ij}$ is the realization for the subject $i$ of the random process $\xproc_j$ and $\varepsilon_{ijk}$ is a measurement error. The measurement errors are supposed i.i.d and following a normal distribution $\mathcal{N}(0,\sigma_j^2)$.

\subsubsection{Process modeling}

As previously stated, considering the set of orthonormal canonical functions $\{f_j^m\}_{1\leq m \leq M}$, each process $j$ from any subject $i$ can be decomposed as:
\begin{equation}
\label{eq:modelblup}
    X_{ij}(t) = \mu_j(t) + \sum_{m=1}^{M} \xi_{ij}^m f^m_j(t)
\end{equation}
where the coefficients $\xi_{ij}^m$ are the basis coefficients associated with the basis $\{f_j^m\}_{1\leq m \leq M}$. Therefore, at the sample level we have:
\begin{equation}
    U_{ijk} = \mu_j(t_{ijk}) + \sum_{m=1}^{M}\xi_{ij}^m f^m_j(t_{ijk}) + \varepsilon_{ijk}
\end{equation}

This formulation allows to see the basis decomposition as a linear mixed-effects model with the fixed-effects part being the mean term and the random-effects part being the decomposition term. Moreover, as $\xi_{ij}^m = \langle X_{ij}, f_j^m \rangle_{L^2}$ and since the deflation strategy introduced in Section \ref{sec:higherorder} leads as previously stated to $\langle \xproc_j, f_j^m \rangle_{L^2} = \langle \xproc_j^m, f_j^m \rangle_{L^2} = \yvec_j^m$ we have that $\xi_{ij}^m = \yvec_{ij}^m$. Therefore, estimating basis coefficients is equivalent to estimating components.

\subsubsection{Estimation}
For simplifying expressions, denoting $N_i = \sum_j n_{ij}$, we write the vector of observations $\bold{U}_i = (\bold{U}_{i1}^{\top}, \dots, \bold{U}_{iJ}^{\top})^{\top}\in\mathbb{R}^{N_{i}\times 1}$, and the mean function vector (at the observation time points) $\boldsymbol{\mu}_i = (\boldsymbol{\mu}_{i,1}^{\top}, \dots, \boldsymbol{\mu}_{i,J}^{\top})^{\top}\in\mathbb{R}^{N_{i}\times 1}$ with $\boldsymbol{\mu}_{i,j} = (\mu_i(t_{ij1}),\dots, \mu_i(t_{ijn_{ij}}))^{\top}\in\mathbb{R}^{n_{ij}\times 1}$. We also write $\bold{F}_{ij}^m=(f_j^m(t_{ij1}), \dots, f_j^m(t_{ijn_{ij}}))^{\top}\in\mathbb{R}^{n_{ij}\times 1}$ and $\bold{F}_{ij} = (\bold{F}_{ij}^1, \dots, \bold{F}_{ij}^M)^{\top}\in\mathbb{R}^{n_{ij}\times M}$, the matrix of the $M$ canonical functions at the observation time points for subject $i$, process $j$, and finally $\boldsymbol{\xi}_{j}=(\xi_j^1, \dots, \xi_j^M)^{\top}\in\mathbb{R}^{M\times 1}$, $\boldsymbol{\xi}=(\boldsymbol{\xi}_1^{\top}, \dots, \boldsymbol{\xi}_J^{\top})^{\top}\in\mathbb{R}^{MJ\times 1}$, the vector of basis coefficients.
Considering that the basis coefficients and the measurement errors are centered and jointly Gaussian, we establish the following proposition, allowing estimating coefficients:
\begin{prop}
Denoting $\bold{F}_i = \diag(\bold{F}_{i1}, \dots, \bold{F}_{iJ})$, $\bold{\Sigma} = \mathbb{E}[\boldsymbol{\xi} \boldsymbol{\xi}^{\top}]$ and $\boldsymbol{\sigma}_{i} = \diag(\sigma_1^2\bold{I}_{n_{i1}}, \dots, \sigma_J^2\bold{I}_{n_{iJ}})$, the best linear unbiaised predictor (BLUP) for $\boldsymbol{\xi}_{i}$ is given by
\begin{equation}
    \mathbb{E}(\boldsymbol{\xi}_i|\boldsymbol{U}_i) = \bold{\Sigma}\bold{F}_i^{\top}(\bold{F}_i\bold{\Sigma}\bold{F}_i^{\top} + \boldsymbol{\sigma}_i)^{-1}(\boldsymbol{U}_i - \boldsymbol{\mu}_i)
\end{equation}
\end{prop}
In this expression, all the terms can be estimated. Notably, the canonical functions can be interpolated at the observation time points of each subject allowing estimating matrices $\bold{F}_{i}$. The noise standard deviation $\sigma_{j}$ can be approximated for each process using the estimated covariance surface of each process (a procedure is presented in \citet{Yao2005}). Finally, the mean functions $\boldsymbol{\mu}_{i}$ are usually estimated when estimating (cross-)covariance surfaces, often using smoothing techniques or GAMs.

\subsection{Retrieving higher-order uncorrelated components}
\label{sec:higherorderuncor}

The deflation strategy introduced in Section \ref{sec:higherorder} allows recovering multiple orthogonal functions for each process. However, as introduced in section \ref{sec:rgcca}, retrieving uncorrelated components is often preferable. In the multivariate setting, this deflation strategy is carried out using the following deflation equation $\xvec_j' = \xvec_j -  (a_j^\top \boldsymbol{\Sigma}_{jj} a_j)^{-1}\boldsymbol{\Sigma}_{jj}a_j a_j^\top\xvec_j$. As previously, we propose to adapt this strategy in the functional setting to deflate the (cross-)covariance operators directly. For this purpose, we establish the following proposition:

\begin{prop}
    Denoting $\boldsymbol{\Sigma}_{jj'}'$ the deflated (cross-)covariance operator of $\boldsymbol{\Sigma}_{jj'}$, obtained from regressing out the components from their associated block. The following equality holds :
    \begin{align}
    \label{defl:type1}
        \boldsymbol{\Sigma}_{jj'}' &= (\boldsymbol{I}_{I_j} - d_j \boldsymbol{\Sigma}_{jj}\boldsymbol{\Phi}_j)\boldsymbol{\Sigma}_{jj'}(\boldsymbol{I}_{I_{j'}} - d_{j'} \boldsymbol{\Phi}_{j'}\boldsymbol{\Sigma}_{j'j'})
    \end{align}
    where $d_j=(\yvec_j^{\top}\yvec_j)^{-1}$ and, as previously, $\boldsymbol{\Phi}_j : L^2(I_j) \rightarrow L^2(I_i)$ is the operator defined by :
    \begin{equation*}
        (\boldsymbol{\Phi}_j)(f) = (f_j \otimes f_j)(f) = \langle f_j, f\rangle f_j
    \end{equation*}
\end{prop}

As previously, new functions associated with uncorrelated components can be obtained by replacing (cross-)covariance operators in the optimization problem by deflated ones and running the solving procedure. However, to obtain uncorrelated estimates of $\yvec_j^m$, additional steps are required. Indeed, the equality $\langle \xproc_j, f_j^m \rangle_{L^2} = \langle \xproc_j^m, f_j^m \rangle_{L^2} = \yvec_j^m$, which was previously used to estimate the components in the mixed-effects model, no longer holds in this setting. Nevertheless, the orthogonal property of the retrieved functions is still holding by construction. Therefore, canonical functions $f_j^m$ can still be used as a decomposition basis in the mixed-effects framework presented previously. Furthermore, using deflation equations, basis coefficients and components can be linked with the following recursive equation:
\begin{equation}
    \yvec_j^{m+1} = \xi_j^{m+1} - \sum_{k=1}^{m} P_j^k \xi_j^{m+1}
\end{equation}
With $P_j^k=({\yvec_j^k}^{\top} \yvec_j^k)^{-1} \yvec_j^k {\yvec_j^k}^{\top}$ being the projection matrix from the regression of $\xi_j^k$ on $\yvec_j^k$, starting with $\xi_j^1 = \yvec_j^1$. This equation can be seen as a decorrelation procedure: for each process, each new basis coefficient estimate is decorrelated from the previous components. 

Finally, the choice of the deflation type depends on the desired usage of the components and the canonical functions. For reconstructing trajectories, orthogonal functions deflation may be more adapted as the orthonormal functions retrieved are preferable for decomposition purposes. For doing clustering or dimension reduction for further analysis, such as regression, using uncorrelated components seem more suitable. 

\subsection{Integrating a multivariate response}

Inspired from the PLS-framework, we propose to modify slightly the FGCCA optimization problem to include a multivariate response $\ymat \in \mathbb{R}^p$ with $p \in \mathbb{N}^*$ for expanding the possibilities given by the framework:
\begin{equation}
\label{eq:fgccacv}
        \argmax_{\substack{f_1, \dots, f_J, \in \Omega_1\times\dots\times\Omega_J\\||a||_2 = 1}} \sum_{j \neq j'} c_{j,j'}g(\langle f_j, \boldsymbol{\Sigma}_{jj'} f_{j'}\rangle_{L^2}) + 2\sum_{j} g(\langle f_j, \boldsymbol{\Sigma}_{j\ymat} a\rangle_{L^2})
\end{equation}
where $\boldsymbol{\Sigma}_{j\ymat}$ is the cross-covariance operator between the process $j$ and the response $\ymat$, defined as:
\begin{align*}
    \boldsymbol{\Sigma}_{j\ymat} &: \mathbb{R}^p \rightarrow L^2(I_j), \; a \mapsto g, \; g(s) = \sum_{i=1}^p \mathbb{E}[\xproc_j(t)\ymat_i] a_i
\end{align*}
As previously, this operator can be estimated with kernel smoothing methods or GAMs.
This additional interaction allows recovering canonical functions for each process that explains best the process, interaction with the others (depending on the design matrix $C$) and interaction with the response. This design is particularly relevant in a predictive framework as it could be interesting to use the components retrieved to predict the response $\ymat$.

Finally, note that the various procedures presented previously are not significantly affected by this change and can be easily rewritten to integrate the multivariate vector using the multivariate-functional cross-covariance operator presented above.

\section{Results}
\label{sec:results}

\subsection{Simulations studies}
\label{sec:simulation}

\subsubsection{Simulation 1: validating the Bayesian approach.} 
\begin{figure}
  \centering
  \includegraphics[scale=0.5]{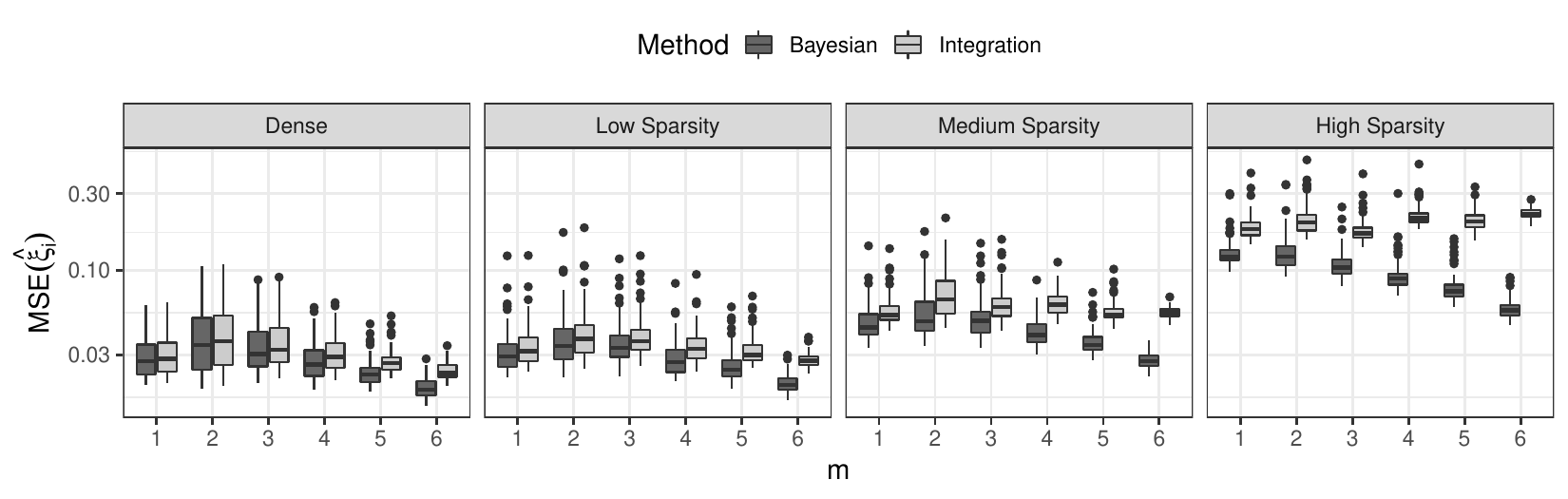}
  \caption{(top) Mean squared error (MSE) boxplots of the components $\yvec_j^m$ estimated with the Bayesian approach or the scalar product (integration) for $m=1,2,3,4,5,6$ obtained from $100$ simulations with $N=100$ and averaged over the $J=3$ processes, $\sigma^2=1$ and various sparsity settings.}
  \label{fig:simulation.design1}
\end{figure}

For $J=3$ processes, we propose to compare the components retrieved using the Bayesian approach previously described in Section (\ref{sec:components}) to the components obtained by computing the scalar product, as usually done. For this purpose, we generate data according to Equation (\ref{eq:modelblup}) so that true component values are known. We choose the first $M=6$ Fourier basis functions in the $[0,1]$ interval as our orthonormal basis. The components $\xi_{ij}^m$ for each subject are generated jointly with a centered Gaussian distribution of covariance $\Sigma$ having a decreasing variance structure. For each subject and each process, the time points are generated by sparsifying a grid of size $50$ in the $[0,1]$ interval. Various sparsity levels are compared: Dense (100 \% of observations retained), Low Sparsity (100 \% to 80 \% of observations retained), Medium Sparsity (80 \% to 40 \% of observations retained), and High Sparsity (40 \% to 10 \% of observations retained). We compare the two approaches by computing the mean squared error on the canonical components for 100 simulations. Results are reported in Figure \ref{fig:simulation.design1}. 

We observe a clear advantage of the Bayesian approach in the Medium and in the High sparsity settings. The gap between the two approaches increases for higher-order components. Furthermore, we notice a slight advantage of the Bayesian approach in the Dense and in the Low sparsity cases, again, especially for the higher-order components. Therefore, we can conclude that the estimation error when using the Bayesian approach is smaller than the estimation error due to the numerical approximation of the integral when using the standard scalar product approach. Thus, we advise using the Bayesian approach in all cases as it is also computationally inexpensive.

\subsubsection{Simulation 2: comparing results} 
\begin{figure}
  \centering
  \includegraphics[scale=0.5]{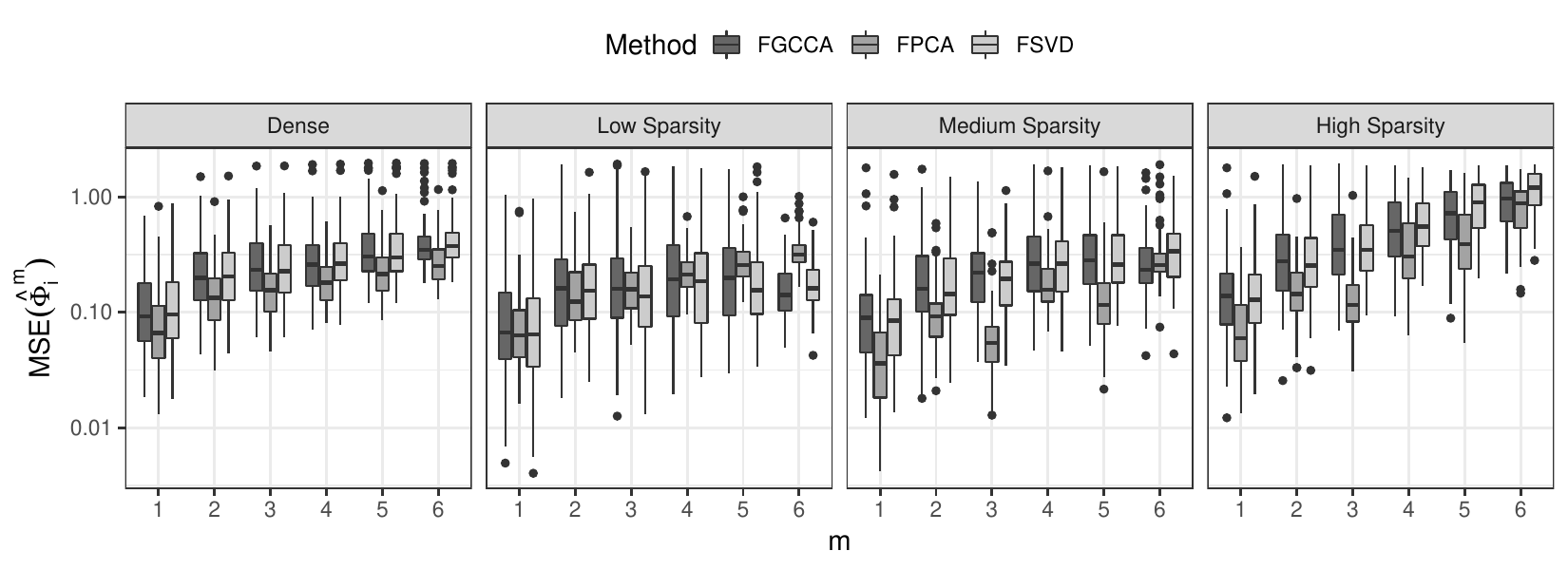}
  \hspace{0.5cm}
  \includegraphics[scale=0.5]{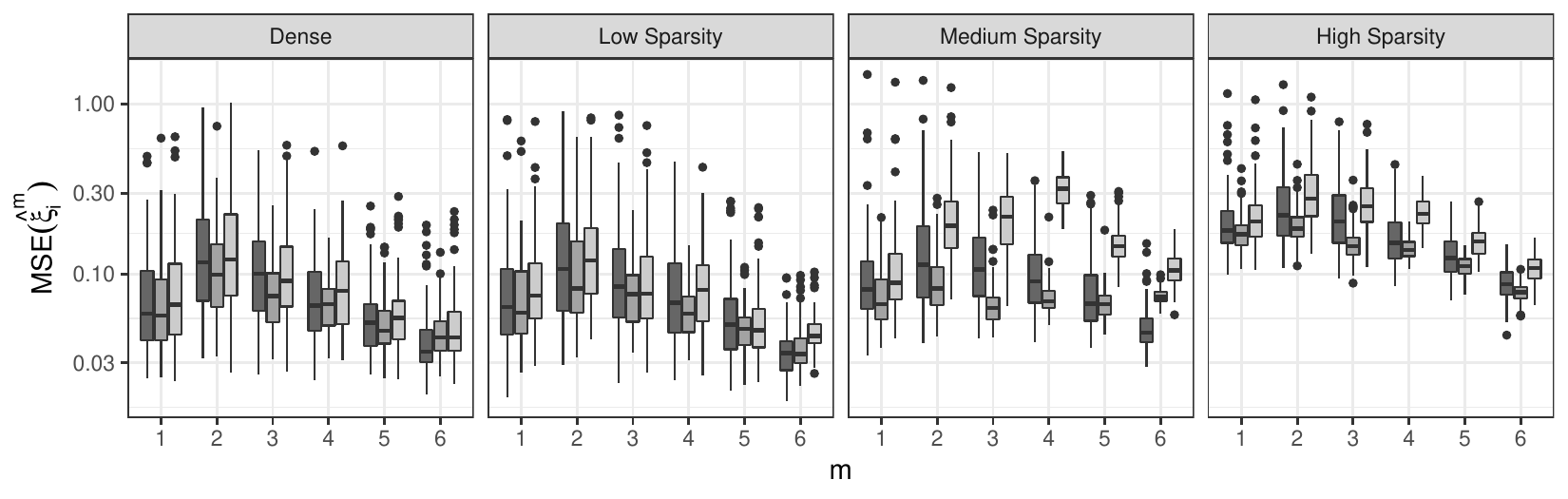}
  \caption{Mean squared errors (MSE) of functions $f^m$ (top) and components $\xi^m$ (bottom) for $m=1,2,3,4,5,6$ obtained from $100$ simulations with $N=100$, $\sigma^2=1$ and various sparsity settings. Comparison between FPCA, FSVD and FGCCA.}
  \label{fig:simulation.design2}
\end{figure}

To validate FGCCA further, we suggest to compare it to some of its subsumed methods. We propose to consider two other approaches for $J=2$ processes: PACE-based Functional Principal Component Analysis (FPCA) (\citet{Yao2005}) and Functional Singular Value Decomposition (FSVD) (\citet{Yang2011}). Both approaches can handle sparse and irregular data. The first gives similar results to FGCCA for specific component's covariance designs. The latter is based on an optimization problem equivalent to FGCCA in the 2 processes setting when regularization parameters $\tau_j$ are set to 1. The simulation setting previously described is used again to generate data. The covariance matrix $\Sigma$ is designed so FPCA, FSVD and FGCCA recover components and functions in the same order. Results are reported on Figure \ref{fig:simulation.design2}.
    
As FGCCA and FSVD are supposed to retrieve similar canonical functions, we observe similar distributions over the mean squared errors for functions. Components estimation rely on a slightly different formula for FSVD. Indeed, the singular values retrieved from the analysis are used, which should, in theory, increase estimation accuracy. However, it seems that in our case, the estimations obtained from a FSVD are significantly worse than with FGCCA, especially for high and medium sparsity settings and for high-order functions. On another hand, FPCA gives slightly better function estimations both for functions and components, especially in the sparse setting and for the first functions and components. For high order components, FGCCA sometimes outperforms FPCA. Additional results along with simulation details are presented in Supplementary Materials. 

\subsubsection{Simulation 3 : comparing reconstruction}

Alternatively, we propose to compare the estimation quality of the reconstructed trajectories between FGCCA with a fully-connected design and an orthogonal deflation, and Multivariate Functional Principal Component Analysis (MFPCA) (\citet{Happ2018}). The reconstructed trajectories are obtained for FGCCA using the decomposition equation (\ref{eq:modelblup}) with the estimated canonical functions and components, and for MFPCA using the multi-dimensional Karhunen-Loeve decomposition, presented in the aforementioned paper. This time, using the package \texttt{funData} (\citet{HappKurz2020}), 3 processes are generated based on the first $M=6$ Fourier basis functions and with a linear decreasing variance over the components.

\begin{figure}
  \centering
  \includegraphics[scale=0.5]{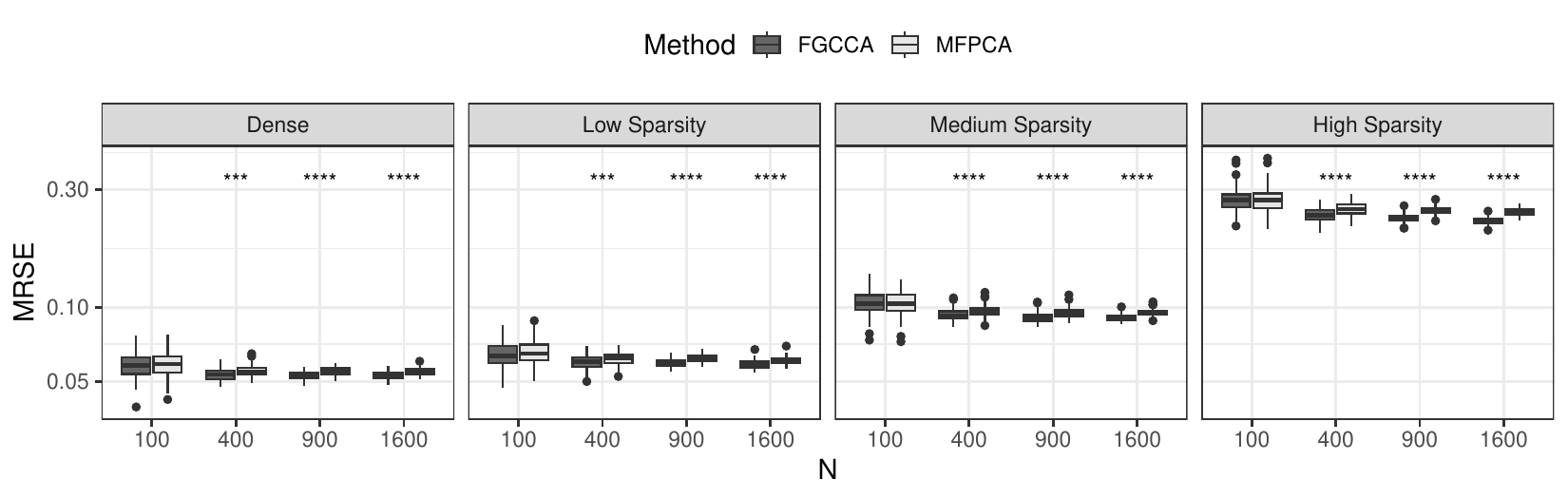}
  \caption{Mean relative squared errors (MRSE) of reconstructed trajectories, using estimated canonical functions and components. Comparing FGCCA and MFPCA with $M=6$, $\sigma^2 = 1$, for various number of subjects $N$ and various sparsity settings. Statistical significance displayed : (***) $p < 0.001$ (****) $p < 0.0001$}
  \label{fig:simulation.design3}
\end{figure}

The mean squared relative error is compared over 100 simulations in various configurations. Results are presented on Figure \ref{fig:simulation.design3}. We can see that our approach improves slightly the reconstruction of the processes, especially when the number of subject $N$ is not too small. The gap between the two methods appears to be stable among all sparsity settings. Further results, available in the Supplementary Materials, show that the difference is bigger as the noise $\sigma^2$ is smaller.

\subsection{Application to \textit{Primary Biliary Cirrhosis} dataset}
\label{sec:application}

\begin{figure}[H]
  \centering
  \includegraphics[scale=0.7]{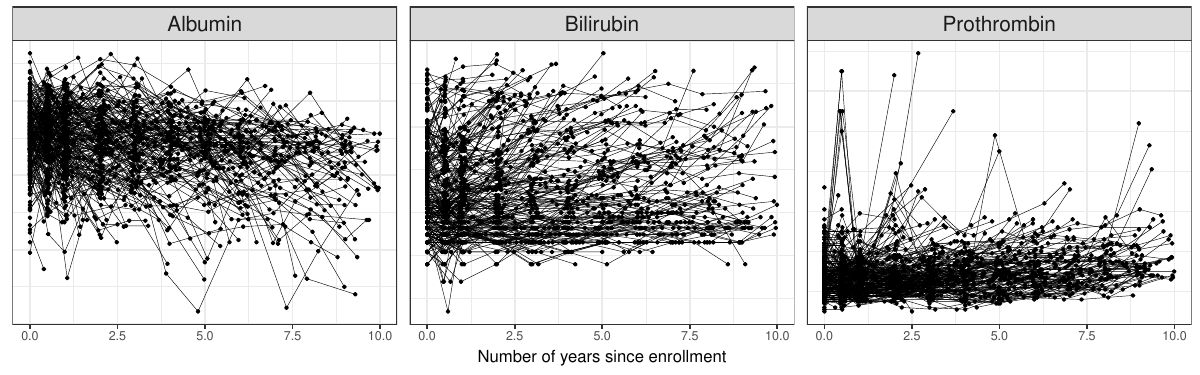}
  \caption{ Longitudinal trajectories for albumin, bilirubin and prothrombin time in the \texttt{pbc2} dataset for all individuals. As usually done, the bilirubin marker is log transformed.}
  \label{fig:datapbc2}
\end{figure}

The Primary Biliary Cirrhosis dataset (\cite{Murtaugh1994}) is a dataset from Mayo Clinic containing the follow-up of various biomarkers extracted from blood analyses of 312 patients who have been diagnosed with primary biliary cirrhosis of the liver, a rare autoimmune disease. We propose to use this multi-biomarker dataset to show various usages of FGCCA. For this purpose, three biomarkers were considered: albumin, bilirubin, which is log transformed, and prothrombin time, observed up to 10 years after the first visit. Those biomarkers were chosen as they have been proven to be good predictors of patient outcomes. Figure \ref{fig:datapbc2} represents an aggregated view of the data.

\subsubsection{Exploratory analysis} 
\begin{figure}
  \centering
  \includegraphics[scale=0.7]{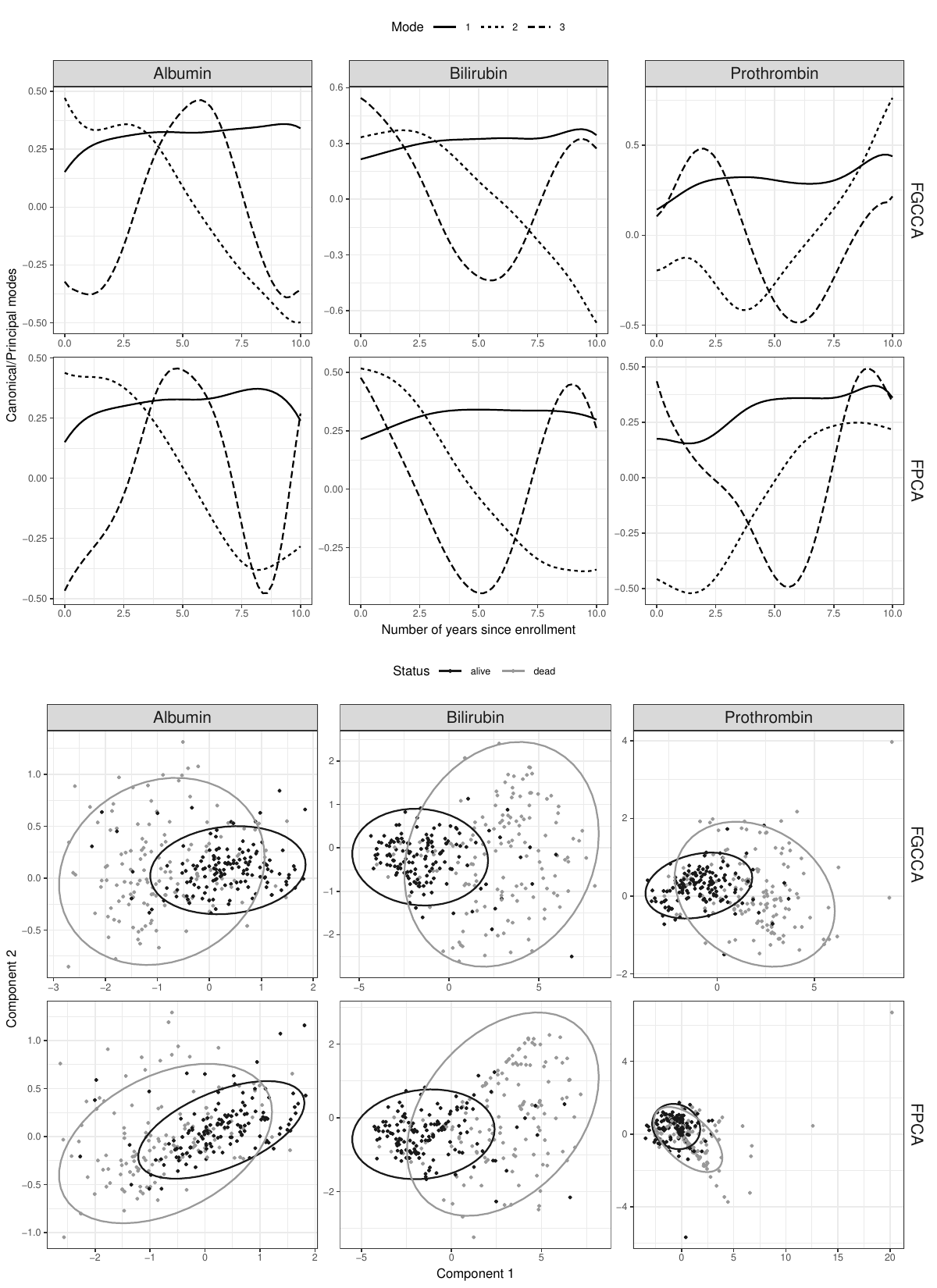}
  \caption{(top) First 3 functional modes retrieved by FGCCA (canonical functions) and FPCA (principal functions), for each biomarker. Functions were flipped to minimize the differences between the two methods. (bottom) Biplots, for each biomarker, of the first 2 components obtained with FGCCA and FPCA coloured by final status. Ellipses represent the estimated Gaussian distributions of the components for each outcome.}
  \label{fig:pbc2.comparison}
\end{figure}

We first propose visualizing the canonical functions and components obtained with FGCCA when using a fully connected design and deflation leading to uncorrelated components. We compare the results to the principal functions and components obtained using PACE-based FPCA (\citet{Yao2005}). For both methods, the bandwidths are manually set to 1 for interpretability. The results are displayed in Figure \ref{fig:pbc2.comparison}. 

The first principal and canonical functions have a similar flat shape, implying that the difference in the trajectories between subjects for all biomarkers comes primarily from an overall shift around the mean. On the other hand, the second canonical and principal functions are either decreasing or increasing during the 10 years interval, indicating that the next source of variation between subjects thus comes from the monotonicity of the trajectories: for the different biomarkers, subjects have either an increasing or decreasing trend. Additionally, we notice a more significant difference between principal and canonical functions for prothrombin, suggesting that this biomarker is particularly correlated to the others. The differences are, however, difficult to interpret. Additionally, note that the functions retrieved with FGCCA are slightly smoother, notably at the end of the interval. We can explain this by the border effects when estimating the (cross-)covariance operators. Indeed, for FGCCA, the functions are estimated using information from multiple processes and not just one (as done in FPCA), leading to more stable and reliable results.

Component plots allow us to see the differences between the two approaches more clearly. First, we notice that the components are spread more evenly for FGCCA, especially for prothrombin, thanks to the normalization. For the three biomarkers, the components given by FGCCA seem to separate better the two outcomes. This property will be confirmed in the predictive analysis.

\subsubsection{Prediction} Inspired by \citet{Singh2019}, we propose using a multiblock functional PLS framework to predict each patient's outcome and demonstrate the ability of FGCCA to integrate a multivariate or univariate response. The multiblock functional PLS design is defined as a FGCCA design where only the associations with the response are considered in the problem. It allows to recover biomarker information correlated to the response, which is particularly useful for predictive purposes. A simple logistic regression model is fitted per biomarker to predict the response using the first component retrieved with FGCCA. For a new subject, the components are predicted using the observed trajectories. The final prediction is obtained by computing a weighted average of the predicted outcomes, where each biomarker prediction is weighted by the correlation of its component with the response. The predictive performances obtained are compared to a similar model where the principal components from FPCA are used instead.

\begin{figure}
  \centering
  \includegraphics[scale=0.7]{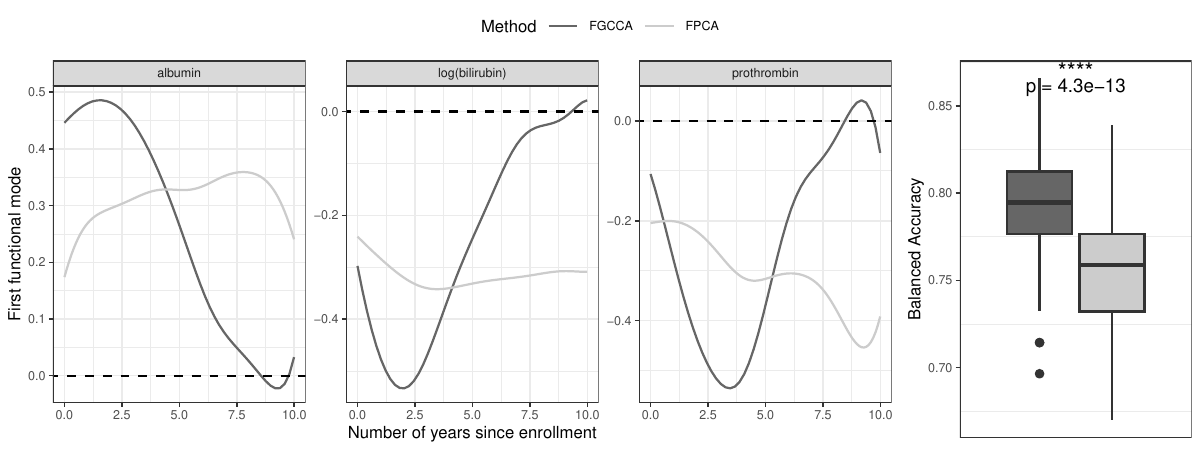}
  \caption{(left) First principal/canonical function retrieved with FPCA/FGCCA. For FGCCA a multiblock-FPLS design is used, integrating only the interaction between each biomarker and the response. Functions are flipped to ensure that the first component is positively correlated to the outcome, to improve interpretability. (right) Boxplot of the balanced accuracy computed on the test set for 100 monte-carlo runs. p-value and significance level of the difference between the two distributions (t-test) are given.}
  \label{fig:pred.comp}
\end{figure}

The results, summarized in Figure \ref{fig:pred.comp}, show that the FGCCA-based components provide a significantly better outcome estimation. The apparent difference in the canonical and principal functions implies that FGCCA has indeed retrieved components highlighting the association with the outcome for each biomarker. As the canonical functions have a monotonic trend, we can argue that survival is mainly associated with a decreasing or increasing behavior of the various biomarkers. More precisely, mortality seems to be associated with an increase in bilirubin, prothrombin and a decrease in albumin. It is consistent with the fact that a decreasing albumin or increasing bilirubin and prothrombin are usually associated with a bad prognosis of the liver.

\subsubsection{Reconstruction} 
\begin{figure}
  \centering
  \includegraphics[scale=0.7]{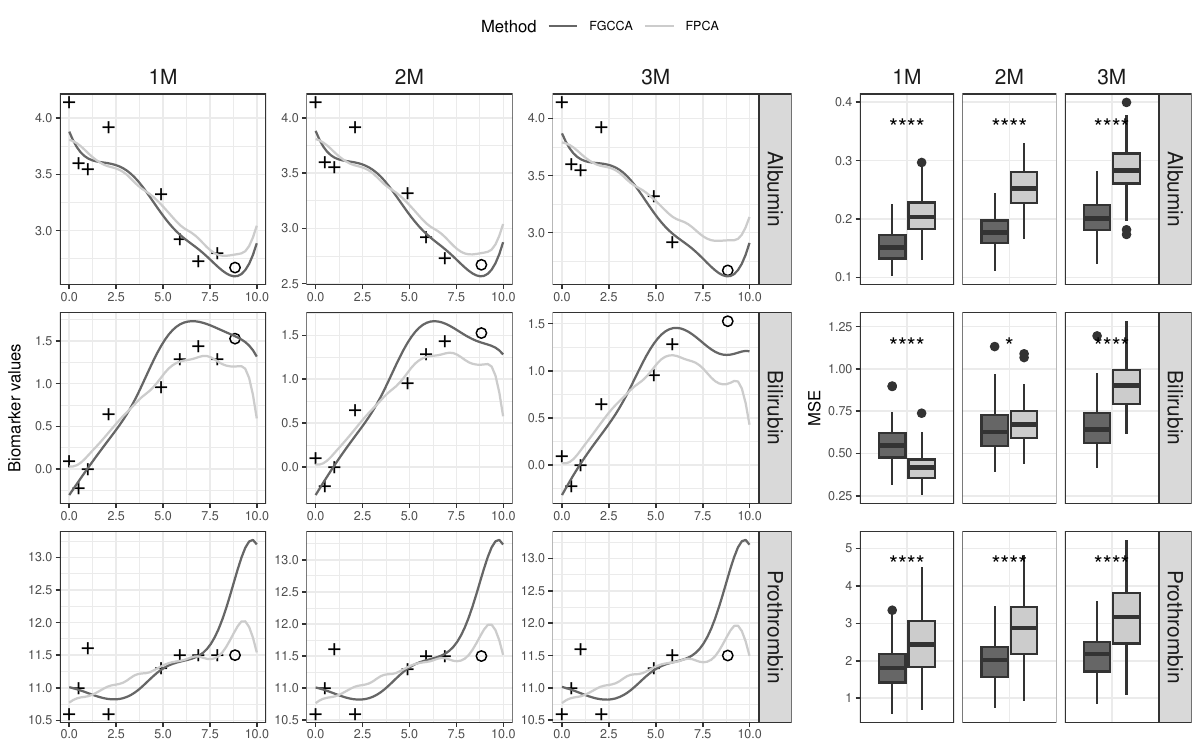}
  \caption{(left) Reconstruction obtained with FGCCA and FPCA in 3 scenarios : (1M) last observation removed, (2M) two last observations removed (3M) three last observations removed. Crosses correspond to observations used to estimate the components, circles correspond to observations we aim to predict. (right) Boxplots of last observation prediction mean squared error obtained on 100 runs. Statistical significance displayed : (*) $p < 0.05$ (****) $p < 0.0001$}
  \label{fig:pbc2.reconstruction}
\end{figure}

Finally, we propose to evaluate the ability of FGCCA to reconstruct trajectories and predict biomarker values at unobserved times. To this end, we propose dividing the data into training and test datasets. The training dataset is used to estimate (cross-)covariance operators and to run FGCCA. The test dataset contains trajectories on which we seek to predict the last observation, which has been removed. Each subject's prediction is made from reconstructed trajectories computed from previously obtained canonical functions (from the training dataset) and components estimated using data before the last observation. We propose to compare the results to an FPCA-based approach, where the components and functions are estimated from an FPCA. Furthermore, to evaluate the robustness of the two approaches, we sparsify the test trajectories at various levels. The results are reported in Figure \ref{fig:pbc2.reconstruction}.

We observe a significant advantage of the FGCCA-based approach over the FPCA-based approach except for bilirubin in the (1M) scenario, asserting that FGCCA has integrated additional or more stable knowledge. We note that the gap widens as the sparsity increases. These results pave the way to joint modeling, as the components could be used both to predict the future trajectories of biomarkers, as it is done here, and the survival of subjects. This promising application is currently investigated.

\section{Discussion}
\label{sec:dicussion}

We introduced Functional Generalized Canonical Correlation Analysis (FGCCA), a flexible framework for exploring associations among multiple longitudinal variables, by finding the main joint modes of variation. The method relies on a monotone and globally convergent algorithm, which only requires (cross-)covariance operators. We proposed a Bayesian approach for estimating the components. Consequently, the method is robust to irregular and sparse data, making it applicable to numerous settings. In addition, we allow integrating a multivariate response in the analysis by slightly modifying the optimization problem, paving the way to mixed data uses. Simulation studies assess the validity of our approach and its underlying design. A wide variety of usages are presented in the application.

As previously mentioned, the method relies significantly on the estimations of (cross-)covariance operators. Therefore, studying more in-depth those estimation procedures could be interesting as new methods have been proposed recently (\cite{Xiao2018}). Computing confidence bands for the estimated scores as it is done in sparse and irregular FPCA (\cite{Yao2005}) is investigated. However, difficulties arise since the FGCCA algorithm does not have a closed-form solution. Finally, an implementation allowing the user to change the regularization parameter was developed. In this context, analyzing the impact of the parameter on the algorithm and the results obtained could be further investigated.

In numerous studies, longitudinal variables can be grouped in blocks representing different modalities. For instance, in imaging genetics, multiple longitudinal variables representing the evolution of several neuroimaging features can be observed along multiple genetic features. In this context, considering a block for each variable, as done with FGCCA, may be inefficient as it would require a complex design and intensive computational resources. Another approach, which is currently being investigated, would be to integrate the multiple longitudinal variables in blocks. This approach was used by \citet{Happ2018} and can be referred to as multivariate functional data modeling. Inspired by the multi-way/tensor literature, a reduced rank model could also significantly help reduce the problem's complexity. In this context, numerous works have been proposed, notably for tensor regression (\citet{Zhou2008}) and, as evoked in the introduction, for RGCCA (\citet{GIRKA}).

\bibliographystyle{abbrvnat}
\bibliography{main}

\appendix

\section{Proofs}

\subsection{Definition of the functional gradient}

Let $\Psi$ be the objective function:
\begin{equation*}
    \Psi(f_1, \dots, f_J) = \Psi(\boldsymbol{f}) = \sum_{j \neq j'} c_{j,j'}g(\langle f_j, \boldsymbol{\Sigma}_{jj'} f_{j'}\rangle_{L^2})
\end{equation*}
Let $j\in \{1,\hdots,J\}$, $f_{j'} \, \forall j'\neq j$.
Let 
$$
\begin{array}{lcl}
\phi_j : L^2( I_j) & \to  & \mathbb{R} \\
         f_j  & \mapsto & \Psi(f_1,\hdots,f_j,\hdots,f_J)\\         
\end{array}
$$
\textbf{Proposition.} $\phi_j$ is Gâteaux-differentiable and its gradient is: 
$$\nabla_j \phi_j (f_j)=
2\sum_{j'\neq j} c_{j, j'} g'(\langle f_j, \boldsymbol{\Sigma}_{jj'}f_{j'}\rangle)\boldsymbol{\Sigma}_{jj'}f_{j'}
$$
\textit{Proof.}
Let $f_j\in L^2(I_j)$. Let's show that it exists a continuous (bounded) linear operator $\phi_j'(f_j)$ such that  
\begin{equation}
 \forall h\in L^2(I_j), \,
\phi_j'(f_j) h 
=\lim_{\substack{\alpha \to 0 \\ \alpha>0}}\dfrac{\phi_j(f_j+\alpha h) - \phi_j(f_j)}{\alpha}  
\label{eq.def_gateaux}
\end{equation}
First notice that according to Fubini theorem, expression of $\phi_j$ may be simplified:
$$\begin{array}{lcl}
\phi_j(f_j)
 & = & 
    \displaystyle \sum_{j' \neq j} c_{j,j'}g(\langle f_j, \boldsymbol{\Sigma}_{jj'} f_{j'}\rangle_{L^2})
            +\sum_{j' \neq j} c_{j',j}g(\langle f_j', \boldsymbol{\Sigma}_{j'j} f_{j}\rangle_{L^2})\\
             & = &  \displaystyle 2\sum_{j' \neq j} c_{j,j'}g(\langle f_j, \boldsymbol{\Sigma}_{jj'} f_{j'}\rangle_{L^2})\\
\end{array}
$$
Then, let's show that the Gâteaux-differential of $\phi_j$ at $f_j$ in the direction $h$ is defined by studying the limit in~(\ref{eq.def_gateaux}):
$$\begin{array}{lcl}
\dfrac{\phi_j(f_j+\alpha h) - \phi_j(f_j)}{\alpha} 
& = & 2\displaystyle\sum_{j' \neq j} c_{j,j'}\dfrac{g(\langle f_j+\alpha h, \boldsymbol{\Sigma}_{jj'} f_{j'}\rangle_{L^2}) - g(\langle f_j, \boldsymbol{\Sigma}_{jj'} f_{j'}\rangle_{L^2})}{\alpha} \\
& = & 2\displaystyle\sum_{j' \neq j} c_{j,j'}
g'(\langle f_j, \boldsymbol{\Sigma}_{jj'} f_{j'}\rangle_{L^2}) 
\langle h, \boldsymbol{\Sigma}_{jj'} f_{j'}\rangle_{L^2}
+o(\alpha)\\
\end{array}  
$$
           Thus, limit in~(\ref{eq.def_gateaux}) exists and leads to a linear operator: 
$$\phi_j'(f_j) h     
=       \left\langle h , 2\displaystyle\sum_{j' \neq j} c_{j,j'}
g'(\langle f_j, \boldsymbol{\Sigma}_{jj'} f_{j'}\rangle_{L^2}) 
\boldsymbol{\Sigma}_{jj'} f_{j'}\right\rangle_{L^2}
 $$
Applying Cauchy-Schwarz and the monotonicity of $g'$, it comes that    $\phi_j'(f_j)$ is bounded (and thus continuous) and so  that $\phi_j(f_j)$ is Gâteaux-differentiable at $f_j$. 
By definition of the gradient: 
      $$\nabla_j \phi_j (f_j)=
2\sum_{j'\neq j} c_{j, j'} g'\left(\langle f_j, \boldsymbol{\Sigma}_{jj'}f_{j'}\rangle_{L^2}\right)\boldsymbol{\Sigma}_{jj'}f_{j'}
$$ 

\subsection{Monotone convergence of the solving procedure (Proposition 1)}
The proof of Proposition 1 is based on the proof given in \citet{Tenenhaus2017}. In the following, we use the notations of the paper. We consider the following lemma:

\begin{lemma}
Consider the set $\Omega = \Omega_1 \times \dots \times \Omega_J$, the function $\Psi : \Omega \rightarrow \mathbb{R}$ and the operator $c : \Omega \rightarrow \Omega$ defined in the paper. The following properties hold:
\begin{enumerate}[label=(\alph*)]
    \item $\Omega$ is a bounded set
    \item $c$ is a continuous operator
    \item $\Psi(\boldsymbol{f}) \leq \Psi(c(\boldsymbol{f}))$ for any $\boldsymbol{f} \in \Omega $
\end{enumerate}
\end{lemma}
\vspace{2mm}
    
\noindent
\textit{Proof of lemma 1} : \\
\\
\textit{(a)} $\Omega$ is bounded as it is the Cartesian product of $J$ bounded sets \\
\\
\textit{(b)} As $\Psi$ is a continuous differentiable operator, $r_j$ is continuous. Thus, $c_j$ is continuous and $c = c_1 \circ \dots c_J$ being the composition of $J$ continuous operators is also continuous\\
\\
\textit{(c)} Let $\ff = ({\ff}_1, \dots, {\ff}_J) \in \Omega$. We want to find an update $\hat{\ff}_j \in \Omega_j$ of ${\ff}_j$ such that $\Psi(\ff) \leq \Psi({\ff}_1, \dots, {\ff}_{j-1}, \hat{\ff}_j, {\ff}_{j+1}, \dots, {\ff}_J)$. For that purpose we use the fact that a convex function lies above its linear approximation. Thus at ${\ff}_j$ and for any $\tilde{\ff}_j \in \Omega_j$ we have
\begin{equation*}
    \Psi({\ff}_1, \dots, {\ff}_{j-1}, \tilde{\ff}_j, {\ff}_{j+1}, \dots, {\ff}_J) \geq \Psi(\ff) + \langle\nabla_j\Psi(\ff), \tilde{\ff}_j - {\ff}_j\rangle = l_j(\tilde{\ff}_j, \ff)
\end{equation*}
Using the Cauchy-Schwartz inequality we obtain the unique maximizer $\hat{\ff}_j \in \Omega_j$ of $l_j(\tilde{\ff}_j, \ff)$ w.r.t. $\tilde{\ff}_j \in \Omega_j$
\begin{equation*}
    \hat{\ff}_j = \argmax_{\tilde{\ff}_j \in \Omega_j} l_j(\tilde{\ff}_j, \ff) = \frac{\nabla_j\Psi(\ff)}{||\nabla_j\Psi(\ff)||} = r_j(\ff)
\end{equation*}
Thus we have,
\begin{equation}
    \label{eq:in1}
    \Psi(\ff) = l_j({\ff}_j, \ff) \leq l_j(r_j(\ff), \ff) \leq \Psi({\ff}_1, \dots, {\ff}_{j-1}, r_j(\ff), {\ff}_{j+1}, \dots, {\ff}_J) = \Psi(c_j(\ff))
\end{equation}
And also,
\begin{equation*}
    \Psi(c_{j-1} \circ \dots \circ c_1(\ff)) \leq \Psi(c_j \circ \dots \circ c_1(\ff))
\end{equation*}
Leading to the desired inequality for any $\ff \in \Omega$:
\begin{equation*}
    \Psi(\ff) \leq \Psi(c_1(\ff)) \leq \Psi(c_2 \circ c_1 (\ff)) \leq \dots \leq \Psi(c_J \circ \dots \circ c_1(\ff)) =  \Psi(c(\ff))
\end{equation*}
\\
Now, using Lemma 1, we can prove Proposition 1. Indeed, Point (c) of Lemma 1 implies that the sequence $\{\Psi(\ff^s)\}$ is monotonically increasing and, since $\Psi$ is bounded on $\Omega$ (Cauchy-Schwartz), it is convergent.\\
\\
\subsection{Deflation equations (Proposition 2 and 4)}

To retrieve orthogonal functions, we project each process $\xproc_j$ onto the orthogonal of the space spanned by its associated normalized canonical function $f_j$. The deflation equation associated to this transformation, giving the transformed process $\xproc_j'$, is:
\begin{equation*}
    \xproc_j' = \xproc_j - \langle f_j, \xproc_j \rangle f_j
\end{equation*}
Denoting $\boldsymbol{F}_j = f_j \otimes f_j$ where $\otimes$ denotes the functional tensor product, defined as $(f \otimes g) (h) = \langle g, h \rangle f$, we can rewrite the previous expression:
\begin{equation*}
 \xproc_j' = ( \boldsymbol{I}_{I_j} - \boldsymbol{F}_{j} ) (\xproc_j)
\end{equation*}
Now, we consider the deflated cross-covariance between deflated processes $\xproc_j'$ and $\xproc_{j'}'$:
\begin{align*}
    (\boldsymbol{\Sigma}_{jj'}'f)(s) &= \int_{I_{j'}} \Sigma_{jj'}'(s, t) f(t) \\
                                    &= \int_{I_{j'}} \mathbb{E}[\xproc_j'(s)\xproc_{j'}'(t)]f(t)\\
                                    &= \mathbb{E}[\int_{I_{j'}} \xproc_j'(s)\xproc_{j'}'(t)f(t)]\\
                                    &= \mathbb{E}[\langle \xproc_{j'}', f \rangle \xproc_j'(s)]
\end{align*}
Allowing us to write :
\begin{align*}
    \boldsymbol{\Sigma}_{jj'}'f &= \mathbb{E}[\langle \xproc_{j'}', f \rangle \xproc_j']\\
                                &= \mathbb{E}[\langle (\boldsymbol{I}_{I_{j'}} - \boldsymbol{F}_{j'})(\xproc_{j'}), f \rangle (\boldsymbol{I}_{I_j} - \boldsymbol{F}_j)(\xproc_j)] \\
                                &= (\boldsymbol{I}_{I_j} - \boldsymbol{F}_j) \mathbb{E}[\langle \xproc_{j'}, (\boldsymbol{I}_{I_{j'}} - \boldsymbol{F}_{j'})(f) \rangle \xproc_j] \\
                                &= (\boldsymbol{I}_{I_j} - \boldsymbol{F}_j) \boldsymbol{\Sigma}_{jj'} (\boldsymbol{I}_{I_{j'}} - \boldsymbol{F}_{j'})(f)
\end{align*}
Leading to desired expression of the cross covariance operators:
\begin{equation*}
    \boldsymbol{\Sigma}_{jj'}' = (\boldsymbol{I}_{I_j} - \boldsymbol{F}_j) \boldsymbol{\Sigma}_{jj'} (\boldsymbol{I}_{I_{j'}} - \boldsymbol{F}_{j'})
\end{equation*}
For retrieving uncorrelated components, $\xproc_j'$ is defined by regressing out from the process the previously retrieved components $\yvec_{j}$. The deflation process can be written as:
\begin{equation}
 \xproc_j' = ( \boldsymbol{I}_{I_j} - d_j \boldsymbol{\Sigma}_{jj} \boldsymbol{F}_{j} ) (\xproc_j)
\end{equation}
where $d_j = \langle f_j, \boldsymbol{\Sigma}_{jj} f_j \rangle^{-1}$. Following the same steps as before we obtain
\begin{equation*}
            \boldsymbol{\Sigma}_{jj'}' = (\boldsymbol{I}_{I_j} - d_j \boldsymbol{\Sigma}_{jj}\boldsymbol{F}_j)\boldsymbol{\Sigma}_{jj'}(\boldsymbol{I}_{I_{j'}} - d_{j'} \boldsymbol{F}_{j'}\boldsymbol{\Sigma}_{j'j'})
\end{equation*}
\subsection{Components conditional expectation (Proposition 3)}

In the following we denote $\boldsymbol{\xi}_{ij} = (\xi_{ij}^1, \dots, \xi_{ij}^M)^{\top}$, $\boldsymbol{\epsilon}_{ij}=(\epsilon_{ij1},\dots, \epsilon_{ijn_{ij}})^{\top}$, $\boldsymbol{\xi}_{i} = (\boldsymbol{\xi}_{i1}^{\top}, \dots, \boldsymbol{\xi}_{iJ}^{\top})^{\top}$ and $\boldsymbol{\epsilon}_{i}=(\boldsymbol{\epsilon}_{i1}^{\top},\dots, \boldsymbol{\epsilon}_{iJ}^{\top})^{\top}$. We propose to write the vector of observations for the process $j$ and subject $i$ with matrix notations:
\begin{equation*}
    \boldsymbol{U}_{ij}
    =
    \boldsymbol{\mu}_{ij} +
    \bold{F}_{ij} \boldsymbol{\xi}_{ij} + 
    \boldsymbol{\epsilon}_{ij}
\end{equation*}
With $\bold{F}_{ij}$ corresponding to a matrix whose columns are the canonical functions at the \textit{observed} time points. Denoting $\bold{F}_i = \diag(\bold{F}_{i1},\dots,\bold{F}_{iJ})$, we also suggest writing the vector of all the observations for the subject $i$ using matrix notations:
\begin{equation*}
    \boldsymbol{U}_{i}
    =
    \boldsymbol{\mu}_{i} +
    \bold{F}_{ij} \boldsymbol{\xi}_{i} + 
    \boldsymbol{\epsilon}_{i}
\end{equation*}
From this, we propose to rewrite jointly the observations and the scores, as:
\begin{equation*}
    \begin{bmatrix}
    \boldsymbol{U}_{i}\\
    \boldsymbol{\xi}_{i}\\
    \end{bmatrix} =
    \begin{bmatrix}
    \boldsymbol{\mu}_{i}\\
    0\\
    \end{bmatrix} 
    +
    \begin{bmatrix}
    \bold{F}_{i} & \bold{I}\\
    \bold{I} & 0
    \end{bmatrix}
    \begin{bmatrix}
    \boldsymbol{\xi}_i\\
    \boldsymbol{\epsilon}_i
    \end{bmatrix}
\end{equation*}
Using this expression, and the jointly Gaussian assumptions on $\xi_i$ and $\epsilon_i$ we clearly see that $\xi_i$ and $\boldsymbol{U}_i$ are also jointly Gaussian with joint law:
\begin{align*}
   \begin{bmatrix}
    \boldsymbol{U}_{i}\\
    \boldsymbol{\xi}_{i}\\
    \end{bmatrix} &\sim
    N \left ( \begin{bmatrix}
    \boldsymbol{\mu}_{i}\\
    0\\
    \end{bmatrix} ,
    \begin{bmatrix}
    \bold{F}_{i} & \bold{I}\\
    \bold{I} & 0
    \end{bmatrix}
    \begin{bmatrix}
     \bold{\Sigma} & 0\\
     0 & \sigma^2 \bold{I} 
    \end{bmatrix}
    \begin{bmatrix}
    \bold{F}_{i}^{\top} & \bold{I}\\
    \bold{I} & 0
    \end{bmatrix}
    \right ) \\
    &\sim
    N \left ( \begin{bmatrix}
    \boldsymbol{\mu}_{i}\\
    0\\
    \end{bmatrix} ,
    \begin{bmatrix}
    \bold{F}_{i} \bold{\Sigma} \bold{F}_i^{\top} + \sigma^2 \bold{I} & \bold{F}_i \bold{\Sigma}\\
    \bold{\Sigma} \bold{F}_i^{\top} & \bold{\Sigma}
    \end{bmatrix}
    \right )
\end{align*}
To obtain the best predictions for the scores, we now consider the conditional distribution of $\xi_i$ and suggest using the conditional expectation as the best predictor. Using the standard formulation of the Gaussian conditional distribution (\cite{Rasmussen2006}) we obtain :
\begin{equation*}
    \mathbb{E}(\boldsymbol{\xi}_i|\boldsymbol{U}_i) = \bold{\Sigma}\bold{F}_i^{\top}(\bold{F}_i\bold{\Sigma}\bold{F}_i^{\top} + \sigma^2\bold{I})^{-1}(\boldsymbol{U}_i - \boldsymbol{\mu}_i)
\end{equation*}

\subsection{Properties of deflated processes and components}

\subsubsection{Orthogonal deflation: correspondence between $\xi_j^m$ and $\yvec_j^m$}

In the decomposition model (\ref{eq:modelblup}), basis coefficients are defined as $\langle \xproc_j, f_j^m \rangle = \xi_j^m$, since $f_j^m$ are orthonormal functions. As in RGCCA, the components are defined in the FGCCA framework as $\yvec_j^m = \langle \xproc_j^m, f_j^m \rangle$ where $\xproc_j^m$ stands for the $m$th deflated processes $j$, with $\xproc_j^0 = \xproc_j$, and $f_j^0$ and $\yvec_j^0$ being respectively the first canonical function and the first component retrieved. We can easily demonstrate recursively that we have for $1\leq m \leq M$:
\begin{equation*}
    \xproc_j^m = \xproc_j - \sum_{k=0}^{m-1} \langle \xproc_j, f_j^k \rangle f_j^k
\end{equation*}
from this we can write
\begin{align*}
    \langle \xproc_j^m, f_j^m \rangle &= \langle \xproc_j, f_j^m \rangle - \sum_{k=0}^{m-1} \langle \xproc_j, f_j^k \rangle \langle f_j^k, f_j^m \rangle \\
    &= \langle \xproc_j, f_j^m \rangle
\end{align*}
since, again, the set of canonical functions $\{ f_j^m \}_{0 \leq m \leq M}$ is a set of orthonormal functions. Therefore, we have proven that $\yvec_j^m = \langle \xproc_j^m, f_j^m \rangle = \langle \xproc_j, f_j^m \rangle = \xi_j^m$

\subsubsection{Uncorrelated deflation: orthogonal property of $f_j^m$}

First, we demonstrate that the functions retrieved using an uncorrelated components deflation strategy are orthogonal. The $(m+1)$th deflation equation for the process $j$ is:
\begin{equation}
 \xproc_j^{m+1} = \xproc_j^m - d_j^m \boldsymbol{\Sigma}_{jj}^m \boldsymbol{F}_{j}^m \xproc_j^m
\end{equation}
where $d_j^m = \langle f_j^m, \boldsymbol{\Sigma}_{jj}^m f_j^m\rangle^{-1} $. We recall that the gradient operator, giving at each iteration the updated block weight function is:
\begin{equation}
\label{eq:grad}
    \langle f_j^{m}, \hat{f}_j^{m+1} \rangle = \nabla_j \Psi(\boldsymbol{f}) = 2\sum_{\substack{j'=1 \\ j'\neq j}}^J c_{j, j'} g'(\langle f_j^{m+1}, \boldsymbol{\Sigma}_{jj'}f_{j'}^{m+1}\rangle)\langle fm_j^{m}, \boldsymbol{\Sigma}_{jj'}^{m+1}f_{j'}^{m+1} \rangle
\end{equation}
Focusing on the elements in the sum we have that
\begin{align*}
    \langle f_j^{m}, \boldsymbol{\Sigma}_{jj'}^{m+1}f_{j'}^{m+1} \rangle &= \langle \mathbb{E}[ \langle \xproc_{j'}^{m+1}, f_{j'}^{m+1} \rangle \xproc_j^{m+1}], f_j^{m} \rangle \\ 
    &= \mathbb{E}[ \langle \xproc_{j'}^{m+1}, f_{j'}^{m+1} \rangle \langle \xproc_j^{m+1}, f_j^{m} \rangle]
\end{align*}
Additionally, 
\begin{align*}
    \langle \xproc_j^{m+1}, f_j^m\rangle &= \langle \xproc_j^{m}, f_j^m\rangle - d_j^m \langle (\boldsymbol{\Sigma}_{jj}^m \bold{F}_j^m)(\xproc_j^m), f_j^m \rangle \\
    &= \langle \xproc_j^{m}, f_j^m\rangle - d_j^m \langle \xproc_j^m, f_j^m \rangle \langle \boldsymbol{\Sigma}_{jj}^m f_j^m, f_j^m \rangle\\
    &= \langle \xproc_j^{m}, f_j^m\rangle - d_j^m \langle \xproc_j^{m}, f_j^m \rangle (d_j^m)^{-1}\\
    &= 0
\end{align*}
Thus we have $\langle f_j^{m}, \boldsymbol{\Sigma}_{jj'}f_{j'}^{m+1} \rangle = 0$ and finally $\langle f_j^m, f_j^{m+1} \rangle = 0$, since all the terms in the sum of the update function are null.\\


\section{Additional experiments}

\subsection{Simulation 2: additional results}

\begin{figure}
  \centering
  \includegraphics[scale=0.5]{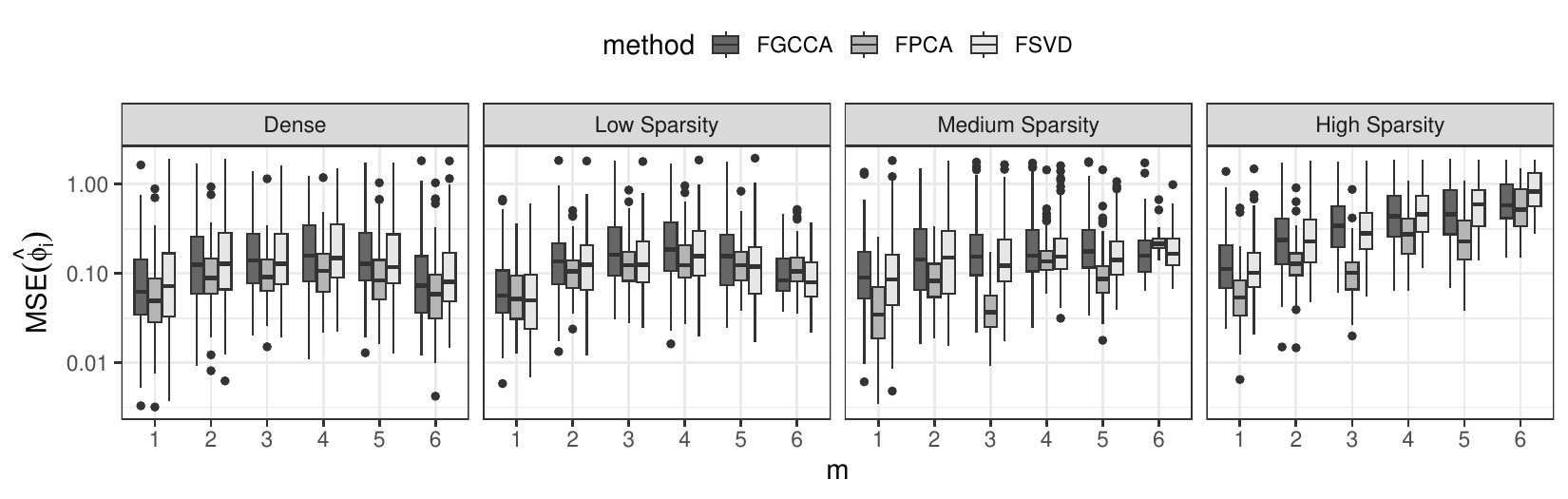}
  \hspace{0.5cm}
  \includegraphics[scale=0.5]{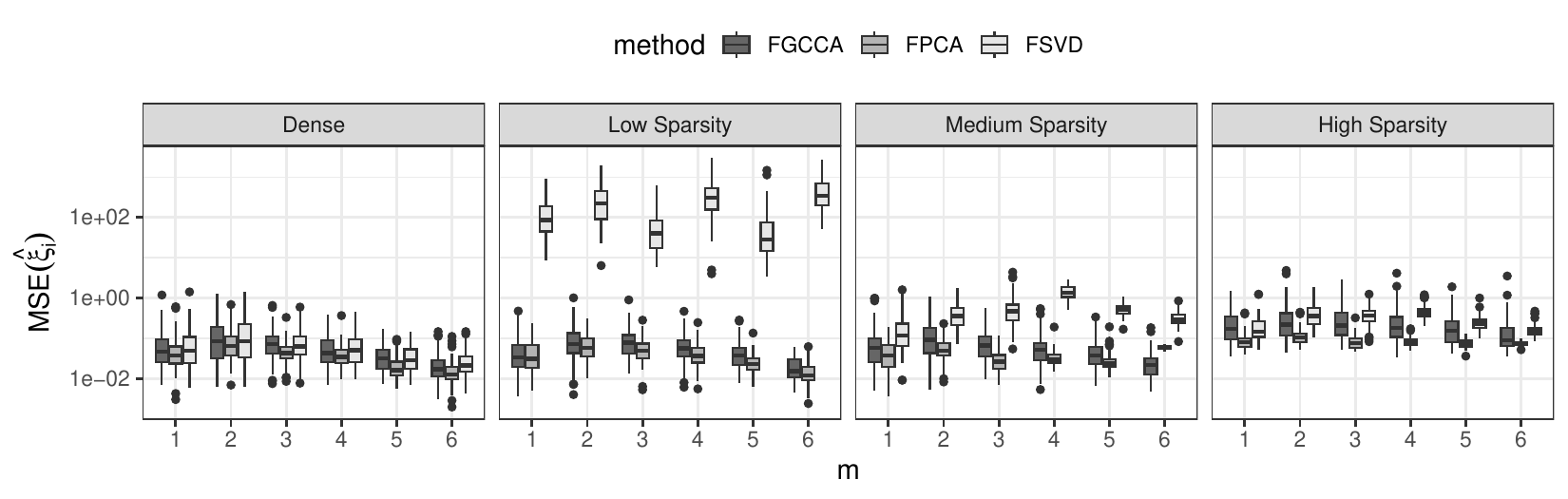}
  \caption{Mean squared errors (MSE) of functions $f_j^m$ (top) and components $\xi_j^m$ (bottom) for $m=1,2,3,4,5,6$ obtained from $100$ simulations with $N=100$, $\sigma^2=0$ and various sparsity settings. Comparison between FPCA (\citet{Yao2005}), FSVD (\citet{Yang2011}) and FGCCA.}
  \label{fig:simulation.design2.1.add}
\end{figure}

\begin{figure}
  \centering
  \includegraphics[scale=0.5]{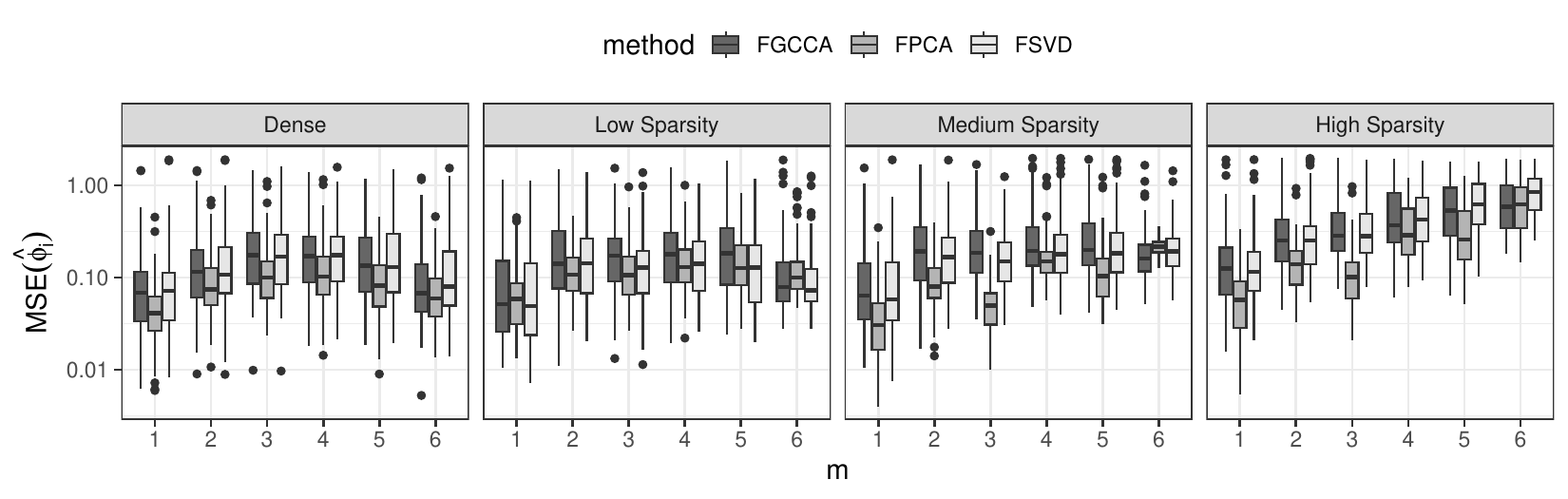}
  \hspace{0.5cm}
  \includegraphics[scale=0.5]{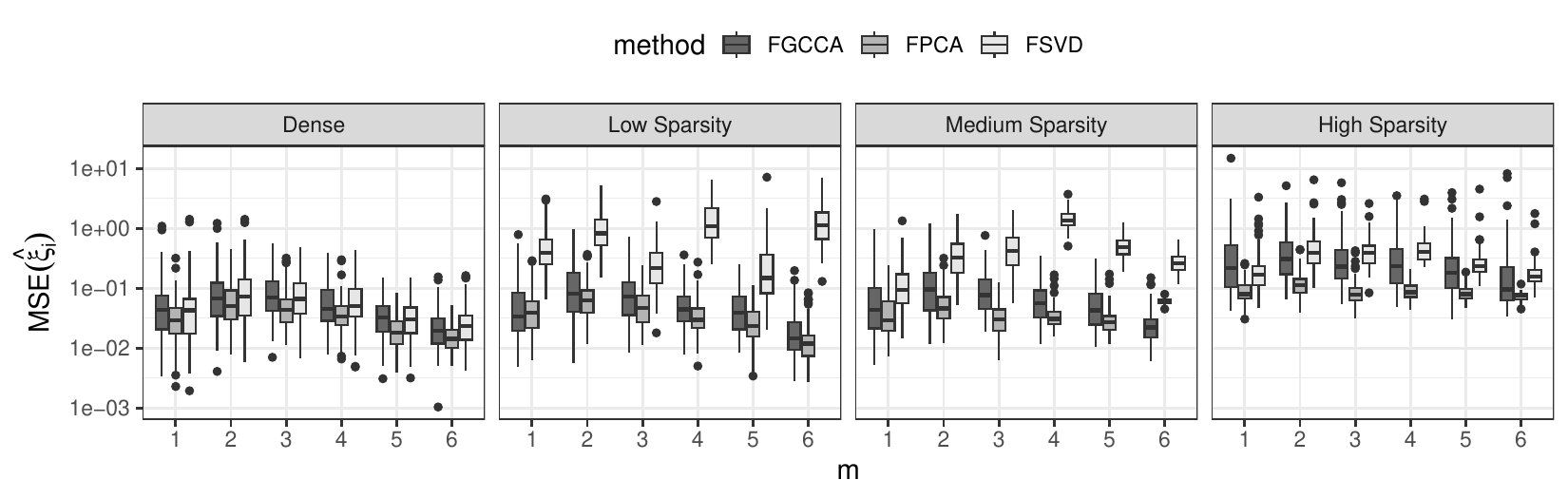}
  \caption{Mean squared errors (MSE) of functions $f_j^m$ (top) and components $\xi_j^m$ (bottom) for $m=1,2,3,4,5,6$ obtained from $100$ simulations with $N=100$, $\sigma^2=0.1$ and various sparsity settings. Comparison between FPCA (\citet{Yao2005}), FSVD (\citet{Yang2011}) and FGCCA.}
  \label{fig:simulation.design2.2.add}
\end{figure}

In this additional section, we investigate the role of $\sigma$ on the estimation accuracy. Results are displayed in figure \ref{fig:simulation.design2.1.add} and \ref{fig:simulation.design2.2.add}.

\subsection{Simulation 3: additional results}

\begin{figure}
  \centering
  \includegraphics[scale=0.5]{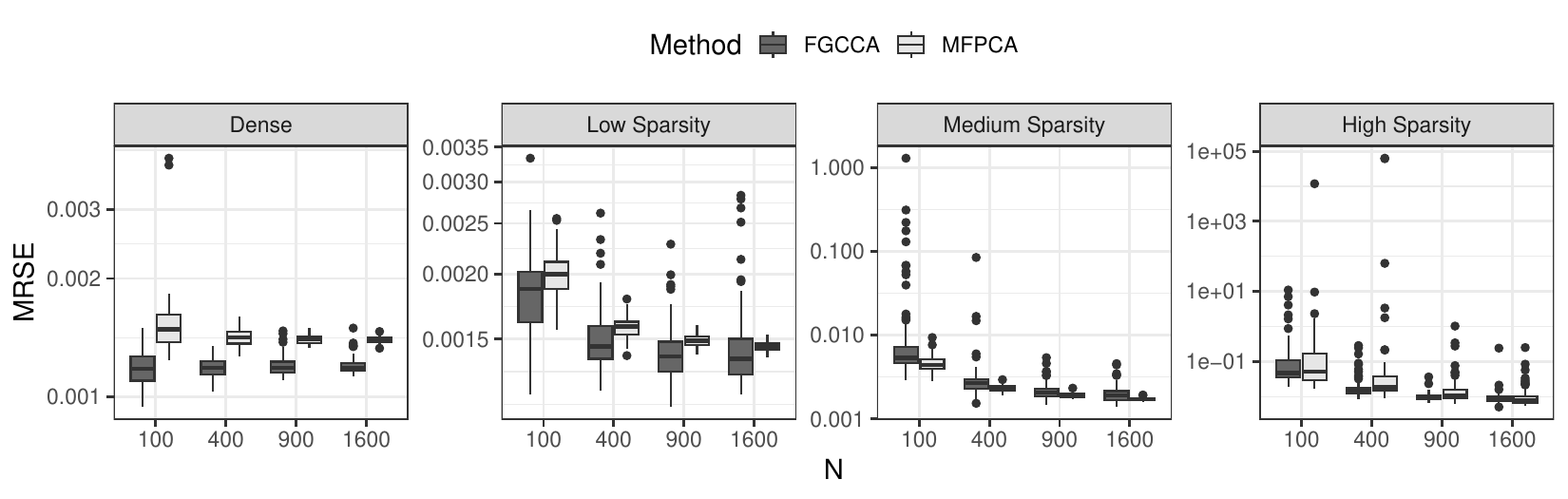}
  \hspace{0.5cm}
  \includegraphics[scale=0.5]{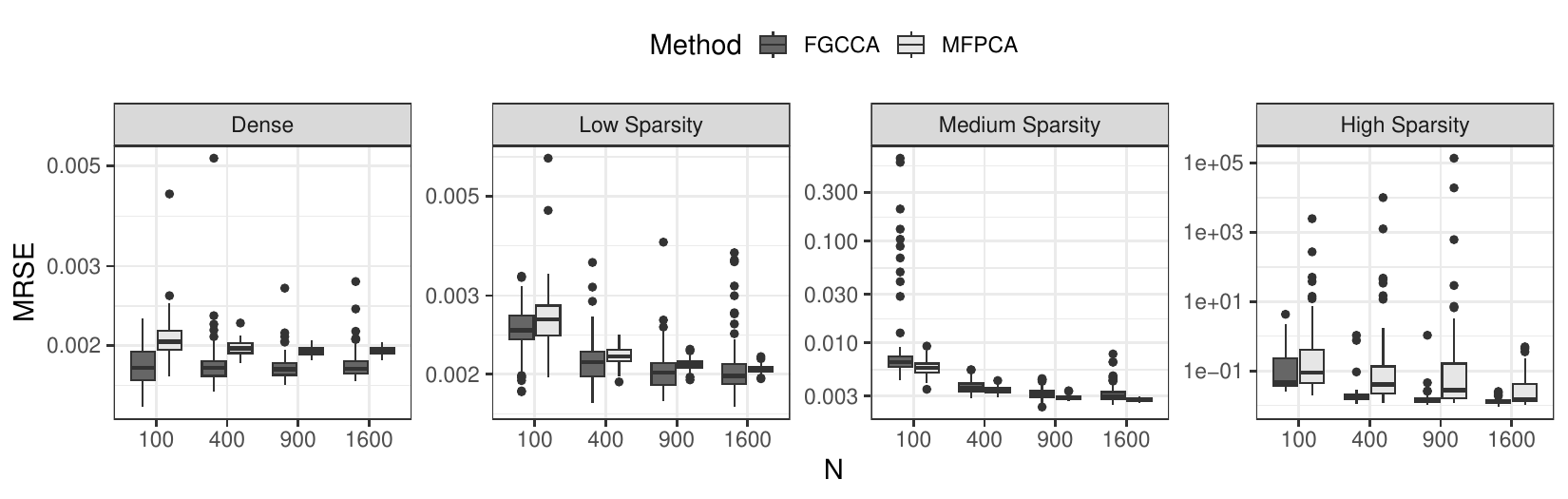}
  \caption{Mean relative squared errors (MRSE) of reconstructed trajectories, using estimated canonical functions and components. Comparing FGCCA and MFPCA with $M=6$, for various number of subjects $N$, various sparsity settings and various noise levels: (top) $\sigma_2 = 0$ and (bottom) $\sigma^2 = 0.1$.}
  \label{fig:simulation.design3.add}
\end{figure}

As previously, we propose additional results obtained with different values of noise $\sigma$. Results are displayed in figure \ref{fig:simulation.design3.add}.

\end{document}